\documentclass[11pt,preprint]{elsarticle}

\usepackage{fancyhdr}
\usepackage{fullpage}
\usepackage{amsmath}
\usepackage{amsbsy}
\usepackage{amssymb}
\usepackage{amscd}
\usepackage{amsfonts}
\usepackage{supertabular}
\usepackage{graphics}
\usepackage{verbatim}
\usepackage{epsfig}
\usepackage{xspace}
\usepackage{euscript}
\usepackage{alltt}
\usepackage{boxedminipage}
\usepackage{float}
\usepackage[colorlinks]{hyperref}
\usepackage{color}
\usepackage[all]{xy}
\usepackage{t1enc}
\usepackage{times,exscale}
\usepackage{graphicx,calc}
\usepackage{subfig}
\usepackage[ruled,vlined]{algorithm2e}
\usepackage{epstopdf}
\usepackage{multirow}
\usepackage{pseudocode}
\usepackage{braket}

\def\bR{{\mathbf{R}}}
\def\bx{{\mathbf{x}}}
\def\bg{{\mathbf{g}}}
\def\R{{\mathbb{R}}}

\journal{arXiv}

\begin{document}

\begin{frontmatter}

\title{SPARC: Accurate and efficient finite-difference formulation and parallel implementation of Density Functional Theory: Isolated clusters}
\author[gatech]{Swarnava Ghosh}
\author[gatech]{Phanish Suryanarayana\corref{cor}}
\address[gatech]{College of Engineering, Georgia Institute of Technology, GA 30332, USA}
\cortext[cor]{Corresponding Author (\it phanish.suryanarayana@ce.gatech.edu) }

\begin{abstract}
As the first component of SPARC (Simulation Package for Ab-initio Real-space Calculations), we present an accurate and efficient finite-difference formulation and parallel implementation of Density Functional Theory (DFT) for isolated clusters. Specifically, utilizing a local reformulation of the electrostatics, the Chebyshev polynomial filtered self-consistent field iteration, and a reformulation of the non-local component of the force, we develop a framework using the finite-difference representation that enables the efficient evaluation of energies and atomic forces to within the desired accuracies in DFT. Through selected examples consisting of a variety of elements, we demonstrate that SPARC obtains exponential convergence in energy and forces with domain size; systematic convergence in the energy and forces with mesh-size to reference plane-wave result at comparably high rates; forces that are consistent with the energy, both free from any noticeable `egg-box' effect; and accurate ground-state properties including equilibrium geometries and vibrational spectra. In addition, for systems consisting up to thousands of electrons, SPARC displays weak and strong parallel scaling behavior that is similar to well-established and optimized plane-wave implementations, but with a significantly reduced prefactor. Overall, SPARC represents an attractive alternative to plane-wave codes for practical DFT simulations of isolated clusters.  

\end{abstract}


\end{frontmatter}

\section{Introduction}
Over the past few decades, the Density Functional Theory (DFT) developed by Hohenberg, Kohn, and Sham \cite{Hohenberg,Kohn1965} has been extensively used for understanding and predicting a wide array of materials properties \cite{jones1989density,ziegler1991approximate,kohn1996density,jones2015density}. The tremendous popularity of DFT---free from any empirical parameters by virtue of its origins in the first principles of quantum mechanics---stems from its high accuracy to cost ratio when compared to other such ab-initio theories \cite{parr1995density,kaduk2011constrained}. However, the efficient solution of the DFT problem still remains a formidable task. In particular, the orthogonality constraint on the Kohn-Sham orbitals in combination with the substantial number of basis functions required per atom results in a cubic scaling with respect to the number of atoms \cite{yang1991direct,carter2008challenges} that is accompanied by a large prefactor. Furthermore, the need for orthogonality gives rise to substantial amount of global communication in parallel computations, which hinders parallel scalability. Consequently, the size of physical systems accessible to DFT has been severely restricted, particularly in the context of ab initio molecular dynamics \cite{marx2009ab,kresse1993ab}, wherein one complete simulation regularly requires the solution of the Kohn-Sham equations tens to hundreds of thousands of times. 

A vast majority of the DFT codes in widespread use today employ plane-waves for discretizing the Kohn-Sham equations \cite{VASP,CASTEP,ABINIT,Espresso,CPMD,DFT++,gygi2008architecture}. The plane-wave basis is an attractive choice because it forms a complete and orthonormal set that is independent of the atomic positions, provides spectral convergence with respect to basis size, and enables the efficient evaluation of convolutions through the Fast Fourier Transform (FFT) \cite{Cooley1965,leszczynski2012handbook}. In addition, effective preconditioners are readily available due to the diagonal representation of the Laplacian operator in this setting \cite{payne1992iterative,hutter1994electronic}. However, the plane-wave basis also suffers from a few notable disadvantages. Specifically, the need for periodic boundary conditions limits its effectiveness in the study of non-periodic and localized systems such as clusters and defects, which typically require the introduction of artificial supercell periodicity \cite{freysoldt2009fully,probert2003improving,Phanish2012}.\footnote{This limitation of plane-waves can be overcome using Hockney's method \cite{Hockney1981}, see for e.g., \cite{Bylaska1996lda}.} Furthermore, the non-locality of plane-waves makes them unsuitable for the development of approaches that scale linearly with respect to the number of atoms \cite{Goedecker,Bowler2012}, and makes parallelization over modern large-scale, distributed-memory computer architectures particularly challenging \cite{bottin2008large,tuckerman2000exploiting}. These characteristics of plane-wave methods are also inherited by the recently developed spectral scheme for isolated clusters \cite{banerjee2015spectral}, which is the analogue of plane-waves in the spherical setting. 

In view of the aforementined limitations, a number of recents efforts have been directed towards the development of real-space DFT implementations. These include discretizations based on finite-differences \cite{chelikowsky1994finite,OCTOPUS,briggs1996real,fattebert1999finite,shimojo2001linear}, finite-elements \cite{Sterne,white1989finite,tsuchida1995electronic,Phanish2010,motamarri2012higher,fang2012kohn,Bylaska2009Adaptive,Batcho1998}, wavelets \cite{arias1999multiresolution,cho1993wavelets,genovese2008daubechies,Harrison2007}, periodic sinc functions \cite{ONETEP}, basis splines (B-splines) \cite{CONQUEST}, non-uniform rational B-splines (NURBS) \cite{masud2012b}, and mesh-free maximum entropy basis functions \cite{Phanish2011}. However, despite the success of real-space methods in overcoming many of the aforementioned limitations---particularly in the flexibility with respect to boundary conditions \cite{souto2015structural}\footnote{Notably, metal-semiconductor interfaces consisting of $\sim 1500$ atoms have been studied (with structural relaxation).}, development of techniques that scale linearly with respect to the number of atoms \cite{ONETEP,CONQUEST}, and scalable high performance computing \cite{hasegawa2011first}\footnote{Simulations of systems consisting of $\sim 100,000$ atoms have been performed, and the Gordon Bell prize has been awarded for this work.} ---plane-wave approaches still remain the preferred choice for practical DFT computations. This is mainly because real-space implementations are unable to consistently outperform the well-optimized plane-wave codes on the modest computational resources commonly available to researchers, while simultaneously achieving the accuracy desired in DFT calculations (Appendix \ref{Appendix:RealSpaceComparison}). Furthermore, the functionality provided by plane-wave codes is significantly larger than their real-space counterparts, having been under development for a longer period of time. 

The finite-difference method is an attractive choice for performing real-space DFT calculations due a number of reasons, including the following. First, the finite-difference discretization results in a standard eigenvalue problem, which can typically be solved more efficiently compared to generalized eigenvalue problems resulting from the use of non-orthogonal bases. Second, the eigenproblem has a relatively small spectral width (i.e., difference between the maximum and minimum eigenvalues), which is critical to the performance of eigensolvers, particularly since effective real-space preconditioners are presently lacking. Third, it is straightforward to employ and switch between high-order approximations, a critical feature for performing efficient and accurate ab-initio calculations. Fourth, the Laplacian has a very compact finite-difference representation, which translates to high computational efficiency. Finally, finite-differences are extremely simple to implement, thereby enabling the rapid prototyping of new solution strategies. These characteristics have motivated the development of DFT packages like PARSEC \cite{chelikowsky1994finite} and OCTOPUS \cite{OCTOPUS}, which now possess most of the features available in mature plane-wave codes (see, e.g. \cite{andrade2015real}). However, the finite-difference method does suffer from a few limitations. The lack of a underlying basis and associated variational structure can result in non-monotonic convergence of the energies and atomic forces. Furthermore, the reduced accuracy of spatial integrations due to the use of a lower order integration scheme can lead to a pronounced `egg-box' effect \cite{ono2010real,BobSchChe15}---phenomenon arising due to the breaking of the translational symmetry---which can significantly affect the accuracy of structural relaxations and molecular dynamics simulations \cite{andrade2015real,li2016atomicbasis,artacho2009}.\footnote{This effect can be diminished by choosing a finer mesh as well as by suitably modifying the pseudopotential \cite{briggs1996real,ono2010real}.} 

In this work, we present an accurate and efficient finite-difference formulation and parallel implementation of DFT for isolated clusters, which forms the first component of SPARC (Simulation Package for Ab-initio Real-space Calculations). The approach employed includes a local reformulation of the electrostatics, the Chebyshev polynomial filtered self-consistent field iteration\footnote{The CheFSI algorithm \cite{zhou2006self} represents a truly significant advance in the context of eigensolvers, and has played a notable role in increasing the efficiency of DFT simulations.},  and a reformulation of the non-local component of the atomic force, which  allows for the efficient evaluation of accurate energies and atomic forces within the finite-difference representation. The electrostatic formulation, atomic force calculation, and overall parallel implementation distinguishes SPARC from existing finite-difference DFT packages like PARSEC \cite{zhou2006parallel} and OCTOPUS \cite{OCTOPUS}.\footnote{A comparison of the scaling and performance of SPARC, PARSEC, and OCTOPUS can be found in Appendix \ref{Appendix:RealSpaceComparison}.} Through a wide variety of examples, we demonstrate that SPARC obtains exponential convergence in energies and forces with domain size; high rates of convergence in the energy and forces to reference plane-wave results on refining the discretization; forces that are consistent with the energy, both being free from any noticeable `egg-box' effect; and accurate ground-state properties (e.g. equilibrium geometries and vibrational spectra). Moreover, SPARC displays similar weak and strong scaling as well-established and optimized plane-wave codes, but with a significantly smaller prefactor. 

The remainder of this paper is organized as follows. In Section \ref{Section:KSDFT}, we provide the mathematical background for DFT. In Section \ref{Sec:FormulationImplementation}, we discuss the finite-difference formulation and efficient parallel implementation of DFT for isolated clusters in SPARC. Next, we verify the accuracy and efficiency of SPARC through selected examples in Section \ref{Sec:Examples}. Finally, we provide concluding remarks in Section \ref{Sec:Conclusions}. 

\section{Density Functional Theory (DFT)} \label{Section:KSDFT}
Consider an isolated system of $N$ atoms comprising of nuclei with valence charges $\{Z_1, Z_2, \ldots, Z_N\}$ and a total of $N_e$ valence electrons. Neglecting spin, the system's free energy in Density Functional Theory (DFT) \cite{Hohenberg,Kohn1965} at finite temperatures \cite{Mermin1965} is of the form\footnote{The free energy is actually a functional of the density matrix rather than the electron density, and is therefore sometimes referred to as Density Matrix Theory.}
\begin{equation} \label{Eqn:Energy}
\mathcal{F} (\Psi, \bg, \bR) =  T_s(\Psi,\bg) + E_{xc}(\rho) +  K(\Psi,\bg,\bR) + E_{el}(\rho,\bR) - TS(\bg) \,,
\end{equation}
where $\Psi = \{ \psi_1, \psi_2, \ldots, \psi_{N_s}\}$ is the collection of orbitals with occupations $\bg = \{g_1, g_2, \ldots, g_{N_s} \}$, $\bR = \{\bR_1, \bR_2, \ldots, \bR_N \}$ is the position of the nuclei, $\rho$ is the electron density, and $T$ is the electronic temperature. The electron density itself depends on the orbitals and their occupations through the relation 
\begin{equation}
\rho(\bx) = 2 \sum_{n=1}^{N_s} g_n \psi_n^2(\bx) \,.
\end{equation}
The first term in Eqn. \ref{Eqn:Energy} denotes the kinetic energy of the non-interacting electrons, the second term corresponds to the exchange-correlation energy, the third term signifies the non-local pseudopotential energy, the fourth term represents the electrostatic energy, and the final term accounts for the contribution of the electronic entropy to the free energy. 

\paragraph{Electronic kinetic energy} In Kohn-Sham DFT, the electronic kinetic energy can be written in terms of the orbitals and their occupations as 
\begin{equation}
T_s(\Psi,\bg) = -\sum_{n=1}^{N_s} g_n \int_{\R^3} \psi_n(\bx) \nabla^2 \psi_n(\bx) \, \mathrm{d \bx} \,.
\end{equation}

\paragraph{Exchange-correlation energy} Since the exact form of the exchange-correlation energy is unknown, a number of approximations have been developed, the  most popular ones being the Local Density Approximation (LDA) \cite{Kohn1965} and the Generalized Gradient Approximation (GGA) \cite{perdew1986accurate}. In this work, we employ the LDA:
\begin{equation}
E_{xc} (\rho) = \int_{\R^3} \varepsilon_{xc} (\rho(\bx)) \rho(\bx) \, \mathrm{d \bx} \,,
\end{equation}
where $\varepsilon_{xc} (\rho) = \varepsilon_x (\rho) + \varepsilon_c (\rho)$ is the sum of the exchange and correlation per particle of a uniform electron gas. 

\paragraph{Non-local pseudopotential energy} The non-local pseudopotential energy can be written as 
\begin{equation}
K(\Psi,\bg,\bR) = 2 \sum_{n=1}^{N_s} g_n \sum_{J=1}^{N} \sum_{lm} \gamma_{Jl} \left( \int_{\R^3} \chi_{Jlm}(\bx,\bR_J) \psi_n(\bx) \mathrm{d \bx} \right)^2 \,,
\end{equation}
where we have employed the Kleinman-Bylander \cite{kleinman1982efficacious} separable form for the pseudopotential. The  coefficients $\gamma_{Jl}$ and projection functions $\chi_{Jlm}$ are of the form
\begin{equation}
\gamma_{Jl} = \left( \int_{\R^3} \chi_{Jlm}(\bx,\bR_J) u_{Jlm}(\bx,\bR_J)  \, \mathrm{d\bx} \right)^{-1} \,, \,\, \chi_{Jlm}(\bx,\bR_J) = u_{Jlm}(\bx,\bR_J) \left(V_{Jl}(\bx,\bR_J)-V_{J}(\bx,\bR_J) \right) \,,
\end{equation}
where $u_{Jlm}$ denote the isolated atom pseudowavefunctions and $V_{Jl}$ represent the angular momentum dependent pseudopotentials, with $l$ and $m$ signifying the azimuthal and magnetic quantum numbers, respectively. In addition, $V_J$ designate the local components of the pseudopotentials, and are typically set to be one of the angular momentum dependent components.  

\paragraph{Electrostatic energy} The electrostatic energy can be further decomposed as 
\begin{equation}
E_{el}(\rho,\bR) = \frac{1}{2} \int_{\R^3} \int_{\R^3} \frac{\rho(\bx)\rho(\bx')}{|\bx - \bx'|} \,\mathrm{d\bx} \, \mathrm{d\bx'} + \sum_{J=1}^{N} \int_{\R^3} \rho(\bx) V_J(\bx,\bR_J)  \, \mathrm{d\bx} + \frac{1}{2} \sum_{I=1}^{N} \sum_{\begin{subarray}{l} J=1 \\J \neq I \end{subarray}}^{N} \frac{Z_{I} Z_{J}}{|\bR_{I}-\bR_{J}|}  \label{Eqn:ElectrostaticEnergy} \,,
\end{equation}
where the first term is the  classical interaction energy of the electron density, also referred to as the Hartree energy. The second term is the interaction energy between the electron density and the nuclei, and the third term is the repulsion energy between the nuclei. 

\paragraph{Electronic entropy} The electronic entropy accounts for the partial orbital occupations, for which we choose the dependence that is appropriate for Fermions:
\begin{equation} 
S(\bg) = -2 k_B \sum_{n=1}^{N_{s}} \left( g_n \log g_n + (1-g_n) \log (1-g_n) \right) \,,
\end{equation}
where $k_B$ is the Boltzmann constant. 

\paragraph{Ground state} The overall ground state in DFT is governed by the variational problem 
\begin{equation} \label{Eqn:GroundStateSplit}
\mathcal{F}_{0}  =   \inf_{\bR} \mathcal{\hat{F}(\bR)}  \,,
\end{equation} 
where
\begin{equation} \label{Eqn:GroundStateElectronic} 
\mathcal{\hat{F}(\bR)} = \inf_{ \Psi ,\bg} \mathcal{F}(\Psi,\bg,\bR) \, \quad s.t. \quad \int_{\R^3} \psi_i(\bx) \psi_j(\bx) \, \mathrm{d\bx} = \delta_{ij} \,, \quad 2 \sum_{n=1}^{N_s} g_n = N_e  \,.
\end{equation}
In this staggered scheme, the electronic ground-state as described by the above equation needs to be computed for every configuration of the nuclei encountered during the geometry optimization represented by Eqn. \ref{Eqn:GroundStateSplit}.  


\section{Formulation and implementation} \label{Sec:FormulationImplementation}
In this section, we describe the real-space formulation and parallel finite-difference implementation of Density Functional Theory (DFT) for isolated clusters. This represents the first component of the first principles code referred to as SPARC, an acronym representing Simulation Package for Ab-initio Real-space Calculations. 

\paragraph{Electrostatic reformulation} The electrostatic energy as presented in Eqn. \ref{Eqn:ElectrostaticEnergy} is inherently non-local, whereby a direct real-space implementation scales as $\mathcal{O}(N^2)$ with respect to the number of atoms. Moreover, it is inefficient in the context of parallel computing since a large amount of interprocessor communication is required. We overcome this by adopting a local formulation of the electrostatics \cite{Pask2005,Suryanarayana2014524}:
\begin{equation} \label{Eqn:ElecEnergyReformulation}   
E_{el}(\rho,\bR) = \sup_{\phi} \bigg \{ - \frac{1}{8 \pi} \int_{\R^3} |\nabla \phi(\bx,\bR)|^2 \, \mathrm{d\bx} + \int_{\R^3}(\rho(\bx)+ b(\bx,\bR)) \phi(\bx,\bR) \, \mathrm{d\bx} \bigg \} - E_{self}(\bR)  + E_c(\bR) \,,    
\end{equation}
where $\phi$ is referred to as the electrostatic potential, and $b$ is the total pseudocharge density of the nuclei. Specifically, 
\begin{equation} 
b(\bx,\bR)= \sum_{J=1}^N b_J(\bx,\bR_J) \,, \quad b_J(\bx,\bR_J)= - \frac{1}{4 \pi} \nabla^2 V_J(\bx,\bR_J) \,, \quad \int_{\R^3} b_J(\bx,\bR_J) \, \mathrm{d\bx} = Z_J \,, \label{Eqn:PseudochargeDefinition} 
\end{equation}
where $b_J$ denotes the pseudocharge density of the $J^{th}$ nucleus that generates the potential $V_J$. The second to last term in Eqn. \ref{Eqn:ElecEnergyReformulation} represents the self energy associated with the pseudocharge densities:
\begin{equation} \label{Eqn:SelfEnergy}
E_{self}(\bR) = \frac{1}{2}\sum_{J=1}^N \int_{\R^3} b_J(\bx,\bR_J) V_J(\bx,\bR_J) \, \mathrm{d\bx} \,. 
\end{equation}
The last term---identically zero for non-overlapping pseudocharge densities---corrects for the error in the repulsive energy when the pseudocharge densities overlap. The explicit expression for $E_c$ can be found in Appendix \ref{Appendix:Correct:RepulsiveEnergy}.

\paragraph{Electronic ground-state} The electronic ground-state for a given position of nuclei is determined by the variational problem in Eqn. \ref{Eqn:GroundStateElectronic}. The corresponding Euler-Lagrange equations are of the form
\begin{eqnarray} 
& & \left( \mathcal{H}\equiv -\frac{1}{2} \nabla^2 + V_{xc} + \phi + V_{nl} \right)  \psi_n  = \lambda_n \psi_n \,, \quad n=1,2, \ldots, N_s \,, \nonumber \\
& &  g_n = \left( 1 + \exp\left( \frac{\lambda_n - \lambda_f}{k_B T} \right) \right)^{-1} \,, \quad \text{where} \,\, \lambda_f \,\,\, \text{is} \,\,\, s.t. \,\,\, 2 \sum_{n=1}^{N_s} g_n = N_e \,, \label{Eqn:EL} \\
& & \rho(\bx) = 2 \sum_{n=1}^{N_s}  g_n \psi_n^2(\bx) \,, \quad -\frac{1}{4 \pi} \nabla^2 \phi(\bx,\bR) = \rho(\bx) + b(\bx,\bR) \,,  \nonumber 
\end{eqnarray}
where $\mathcal{H}$ is the Hamiltonian operator, $V_{xc}= \delta E_{xc}/\delta \rho$ is the exchange-correlation potential, 
\begin{equation}
V_{nl}f = \sum_{J=1}^N V_{nl,J}f = \sum_{J=1}^{N} \sum_{lm} \gamma_{Jl} \chi_{Jlm} \int_{\R^3} \chi_{Jlm}(\bx,\bR_J) f(\bx) \, \mathrm{d\bx}
\end{equation}
is the non-local pseudopotential operator, and $\lambda_f$ is the Fermi energy. 

The electronic ground-state is determined using the Self-Consistent Field (SCF) method \cite{slater1974self}. Specifically, the non-linear eigenvalue problem described in Eqn. \ref{Eqn:EL} is solved using a fixed-point iteration---accelerated using mixing/extrapolation schemes \cite{fang2009two,lin2013elliptic,pratapa2015restarted,banerjee2015periodic}---with respect to the potential $V_{eff} = V_{xc} + \phi$. In each iteration of the SCF method, the electron density is calculated by solving for the eigenfunctions of the linearized Hamiltonian, and the effective potential is evaluated by solving the Poisson equation for the electrostatic potential. Indeed, the calculation of the orthonormal Kohn-Sham orbitals scales asymptotically as $\mathcal{O}(N^3)$ with respect to the number of atoms. In order to overcome this restrictive scaling, $\mathcal{O}(N)$ approaches \cite{Goedecker,Bowler2012} will be subsequently developed and implemented into SPARC.

\paragraph{Free energy} In SPARC, the free energy is evaluated using the Harris-Foulkes \cite{harris1985simplified,foulkes1989tight} type functional: 
\begin{eqnarray} 
\mathcal{\hat{F}(\bR)} & = & 2\sum_{n=1}^{N_s} g_n \lambda_n + \int_{\R^3} \varepsilon_{xc} (\rho(\bx)) \rho(\bx) \, \mathrm{d \bx} - \int_{\R^3} V_{xc}(\rho(\bx))\rho(\bx)\, \mathrm{d\bx}  + \frac{1}{2} \int_{\R^3}(b(\bx,\bR)-\rho(\bx)) \phi(\bx,\bR) \, \mathrm{d\bx} \nonumber \\
& - & E_{self}(\bR)  + E_c(\bR) + 2 k_B T \sum_{n=1}^{N_{s}} \left( g_n \log g_n + (1-g_n) \log (1-g_n) \right) \,, \label{Eqn:FreeEnergy:GroundState:DFT} 
\end{eqnarray}
where $E_{self}$ and $E_c$ are as defined in Eqns. \ref{Eqn:SelfEnergy} and \ref{Eqn:RepulsiveCorrectionEnergy}, respectively. 

\paragraph{Atomic forces} Once the electronic ground-state has been determined, the atomic forces are calculated using the following expression: 
\begin{eqnarray}
\mathbf{f}_J & = & -\frac{\partial \mathcal{\hat{F}}(\bR) }{\partial \bR_J} \nonumber \\ 
 & = & \int_{\R^3} \nabla b_{J}(\bx,\bR_{J}) \left(\phi(\bx,\bR)-V_{J}(\bx,\bR_{J})\right) \, \mathrm{d\bx} + \mathbf{f}_{J,c}(\bR)  \label{Eqn:Force:Nuclei} \\
 & & -4\sum_{n=1}^{N_s} g_n \sum_{lm} \gamma_{Jl} \left( \int_{\R^3} \psi_n(\bx) \chi_{Jlm}(\bx,\bR_J) \, \mathrm{d\bx} \right) \left( \int_{\R^3} \nabla \psi_n(\bx) \chi_{Jlm}(\bx,\bR_J) \, \mathrm{d\bx} \right) \nonumber  \,.
\end{eqnarray}
The first term is the local component of the force \cite{Phanish2012}, and the second term---expression presented in Appendix \ref{Appendix:Correct:RepulsiveEnergy}---represents the electrostatic correction in the forces when the pseudocharge densities overlap \cite{Suryanarayana2014524}. The final term, which represents the non-local component of the atomic force, has been obtained by transferring the derivative on the non-local projectors (with respect to the atomic position) to the orbitals (with respect to space) \cite{hirose2005first}. This strategy has been adopted since the orbitals are typically much smoother than the projectors, which enables more accurate atomic forces to be obtained \cite{pratapa2015spectral}. 

\begin{figure}[H]
\centering
\includegraphics[keepaspectratio=true,width=0.71\textwidth]{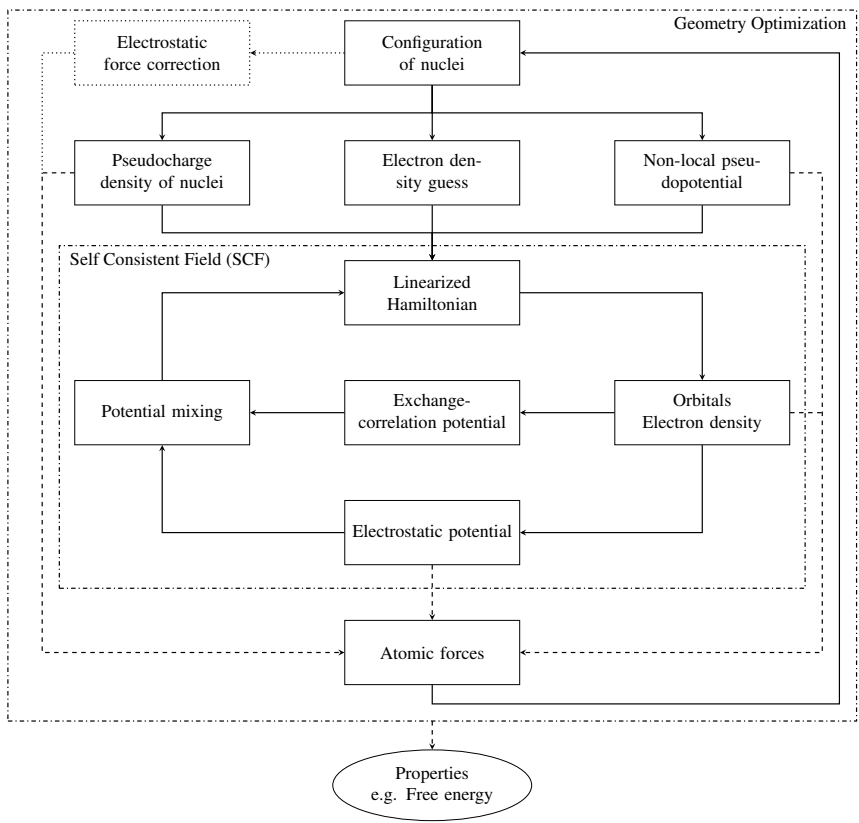} 
\caption{Outline of ground-state DFT simulations in SPARC.}
\label{Fig:flowchart}
\end{figure}

\paragraph{Overview of SPARC} SPARC has been implemented in the framework of the Portable, Extensible Toolkit for scientific computations (PETSc) \cite{Petsc1,Petsc2} suite of data structures and routines. The electronic and structural ground-states for isolated clusters are determined using the methodology outlined in Fig. \ref{Fig:flowchart}, whose key components are discussed in detail in the subsections below.

\subsection{Finite-difference discretization} \label{Sec:FD}
The simulations are performed on a cuboidal domain $\Omega$ with boundary $\partial \Omega$ and sides of length $L_1$, $L_2$ and $L_3$. The domain $\Omega$ is discretized using a uniform finite-difference grid with spacing $h$ such that $L_1=n_1 h$, $L_2=n_2h$ and $L_3=n_3h$, where $n_1,n_2,n_3 \in \mathbb{N}$, $\mathbb{N}$ being the set of all natural numbers. Each node in the finite-difference grid is indexed by $(i,j,k)$, where $i=1,2,\ldots, n_1$, $j=1,2,\ldots, n_2$ and $k=1,2,\ldots, n_3$. We approximate the Laplacian of any function $f$ at the grid point $(i,j,k)$ using finite-differences:
\begin{eqnarray}\label{Eqn:FD:Laplacian}
\nabla^2_h f \big|^{(i,j,k)} \approx \sum_{p=0}^{n_o} w_p \bigg(f^{(i+p,j,k)} + f^{(i-p,j,k)} + f^{(i,j+p,k)} + f^{(i,j-p,k)} + f^{(i,j,k+p)} + f^{(i,j,k-p)} \bigg) \,,
\end{eqnarray}
where $f^{(i,j,k)}$ represents the value of the function $f$ at the node $(i,j,k)$. The weights $w_p$ are given by \cite{mazziotti1999spectral,ghosh2014higher}
\begin{eqnarray}
w_0 & = & - \frac{1}{h^2} \sum_{q=1}^{n_o} \frac{1}{q^2} \,, \nonumber \\
w_p & = & \frac{2 (-1)^{p+1}}{h^2 p^2} \frac{(n_o!)^2}{(n_o-p)! (n_o+p)!} \,, \,\, p=1, 2, \ldots, n_o. \label{Eqn:weightsLap}
\end{eqnarray} 
Similarly, we approximate the gradient using the finite-difference approximation:\footnote{In this work, we employ the finite-difference gradient operator for the calculation of the atomic forces. Once sophisticated exchange-correlation functionals (e.g. GGA \cite{perdew1986accurate}) are incorporated into SPARC, we will approximate the gradient of the electron density in similar fashion.}
\begin{eqnarray}\label{Eqn:gradient:approximate}
\nabla_h f \big|^{(i,j,k)} \approx \sum_{p=1}^{n_o} \tilde{w}_p \bigg( ( f^{(i+p,j,k)} - f^{(i-p,j,k)}) \hat{\mathbf{e}}_1 + ( f^{(i,j+p,k)} - f^{(i,j-p,k)}) \hat{\mathbf{e}}_2  + ( f^{(i,j,k+p)} - f^{(i,j,k-p)}) \hat{\mathbf{e}}_3 \bigg) \,,
\end{eqnarray}
where $\hat{\mathbf{e}}_1$, $\hat{\mathbf{e}}_2$ and $\hat{\mathbf{e}}_3$ signify unit vectors along the edges of $\Omega$, and the weights \cite{mazziotti1999spectral,ghosh2014higher}
\begin{equation}
\tilde{w}_p = \frac{(-1)^{p+1}}{h p} \frac{(n_o!)^2}{(n_o-p)! (n_o+p)!} \,, \,\, p=1, 2, \ldots, n_o.
\end{equation}
These finite-difference expressions for the Laplacian and gradient represent $\mathcal{O}(h^{2n_o})$ accurate approximations. We enforce zero Dirichlet boundary conditions by setting $f^{(i,j,k)}=0$ for any index that does not correspond to a node in the finite-difference grid. While performing spatial integrations, we assume that the function $f$ is constant in a cube of side $h$ around each grid point, i.e.,\footnote{Even though the derivatives are approximated using high-order finite-differences, we approximate the integrals using the low-order midpoint integration rule. In doing so, the discrete free energy obtained is consistent with the discrete Kohn-Sham equations, i.e., the calculated electronic ground-state corresponds to the minimum of the free energy within the finite-difference approximation.}
\begin{equation} \label{Eqn:IntApprox}
\int_{\Omega} f(\bx) \, \mathrm{d\bx} \approx h^3  \sum_{i=1}^{n_1} \sum_{j=1}^{n_2} \sum_{k=1}^{n_3}f^{(i,j,k)} .
\end{equation} 
Using this integration rule, we approximate the non-local pseudopotential operator as 
\begin{equation} \label{Eqn:non-localPS:FD}
V_{nl} f \big|^{(i,j,k)} = \sum_{J=1}^{N} V_{nl,J} f \big|^{(i,j,k)} \approx h^3 \sum_{J=1}^N \sum_{lm} \sum_{p=1}^{n_1} \sum_{q=1}^{n_2} \sum_{r=1}^{n_3} \gamma_{Jl} \chi_{Jlm}^{(i,j,k)} \chi_{Jlm}^{(p,q,r)} f^{(p,q,r)} \,.
\end{equation}
 
Henceforth, we denote the Hamiltonian matrix resulting from the above discretization by $\mathbf{H} \in \R^{N_d \times N_d}$, where $N_d=n_1 \times n_2 \times n_3$ is the total number of finite-difference nodes used to discretize $\Omega$. In addition, we represent the eigenvalues of $\mathbf{H}$ arranged in ascending order by $\lambda_1, \lambda_2, \ldots, \lambda_{N_d}$. We store the discrete Laplacian in compressed row format, apply the nonlocal pseudopotential in a matrix-free way, and store the discrete orbitals as the columns of the dense matrix ${\bf \Psi} \in \R^{N_d \times N_s} $. During parallel computations, we partition the domain as $\Omega = \bigcup\limits_{p=1}^{n_p} \Omega_p$, where $\Omega_p$ denotes the domain local to the $p^{th}$ processor, and $n_p$ is the total number of processors. The specific choice of $\Omega_p$ corresponds to the PETSc  default for structured grids. 

\subsection{Pseudocharge density generation and self energy calculation} \label{Subsec:PseudochargeSelfEnergy}
In each step of geometry optimization, the pseudocharge densities are assigned to the grid using the finite-difference approximated Laplacian \cite{Phanish2012,Suryanarayana2014524}:
\begin{equation} \label{Eqn:PseudochargeDiscrete}
b^{(i,j,k)}= \sum_{J=1}^N b_J^{(i,j,k)} \,, \quad b_J^{(i,j,k)}= - \frac{1}{4 \pi} \nabla^2_h V_J\big|^{(i,j,k)} \,.
\end{equation}
The associated discrete self energy is of the form
\begin{equation} \label{Eqn:DiscreteEself}
E_{self}^h = \frac{1}{2} h^3 \sum_{J=1}^N \sum_{i=1}^{n_1} \sum_{j=1}^{n_2} \sum_{k=1}^{n_3} b_J^{(i,j,k)} V_J^{(i,j,k)} \,.
\end{equation}
Since each radially symmetric pseudopotential $V_J$ matches the Coulomb potential outside some prespecified cutoff radius $r_J^c$, the continuous pseudocharge density $b_J$ has compact support in a sphere of radius $r_J^c$ centered at $\bR_J$. This is not the case for the corresponding discrete pseudocharge density $b_J^{(i,j,k)}$, which actually has infinite extent due to the use of the finite-difference Laplacian (Eqn. \ref{Eqn:PseudochargeDiscrete}). However, $b_J^{(i,j,k)}$ has exponential decay away from $\mathbf{R}_J$ (Appendix \ref{Appendix:Pseudocharge}), which allows for truncation at some suitably chosen radius $r_J^b$. It is worth noting that even though the discrete pseudocharge densities may overlap, as long as there is no overlap between the continuous pseudocharge densities, the electrostatic correction to the energy and forces (i.e., $E_c$ and $\mathbf{f}_{J,c}$) both rapidly converge to zero as the mesh is refined. This is a consequence of the finite-difference Laplacian being used to assign the pseudocharge densities on to the mesh, with the corresponding inverse operation being performed during the solution of the Poisson equation in Eqn. \ref{Eqn:EL}. It is also worth noting that even though the pseudopotential $V_J$ might not be smooth---particularly at the cutoff radius $r_J^c$---we employ a higher-order finite-differences for generating the pseudocharges (Eqn. \ref{Eqn:PseudochargeDiscrete}), since they result in smaller values of $r_J^b$ ($>r_J^c$), i.e., decay of $b_J$ is faster due to a better approximation of the Laplacian.

We calculate the total pseudocharge density $b^{(i,j,k)}$ and the corresponding self energy $E_{self}^h$ using the approach outlined in Algorithm \ref{Algo:PseudochargeCalculation}. We use $P_{r_J^b}^p$ to denote the set of all atoms whose $\Omega_{r_J^b}$---cube with side of length $2r_J^b$ centered on the $J^{th}$ atom---has overlap with the processor domain $\Omega_p$. We have chosen a cube rather than a sphere due to its simplicity and efficiency within the Euclidean finite-difference discretization. The value of $r_J^b$ for every type of atom---determined at the start of the complete DFT simulation---is chosen such that the charge constraint in Eqn. \ref{Eqn:PseudochargeDefinition} is satisfied to within a prespecified normalized tolerance $\varepsilon_b$, i.e.,
\begin{equation} \label{Eqn:PseudochargeRadiusChoice}
\left| \frac{h^3 \sum_{i=1}^{n_1} \sum_{j=1}^{n_2} \sum_{k=1}^{n_3} b_J^{(i,j,k)} - Z_J}{Z_J} \right| <  \varepsilon_b \,. 
\end{equation} 
While describing Algorithm \ref{Algo:PseudochargeCalculation}, we use the subscripts $s$ and $e$ to denote the starting and ending indices of $\Omega_{r_J^b} \cap \Omega_p \neq \emptyset$, respectively. In this overlap region (and an additional $2 n_0$ points in each direction), we interpolate $V_J^{(i,j,k)}$ on to the finite-difference grid using cubic-splines \cite{ahlberg1967theory}. Next, we utilize Eqns. \ref{Eqn:PseudochargeDiscrete} and \ref{Eqn:DiscreteEself} to compute $b^{(i,j,k)}$ and $E_{self}^{h,p}$, where $E_{self}^{h,p}$ is the contribution of the $p^{th}$ processor to the self energy. Finally, we sum the contributions from all the processors to obtain the total self energy $E_{self}^h$. 

The local and independent nature of the aforedescribed computations  ensure that they possess good weak and strong parallel scalability. In addition, they scale as $\mathcal{O}(N)$ with respect to the number of atoms, which makes them efficient even for large systems. In order to achieve perfect $\mathcal{O}(N)$ scaling in practice, the atoms need to be suitably distributed amongst the processors in $\mathcal{O}(N)$ time, after which each processor is only required to go over the local subset of atoms. However, we determine $P_{r_J^b}^p$ by going over all the atoms, which makes the overall procedure formally slightly worse than $\mathcal{O}(N)$ with respect to the number of atoms. Since the system sizes studied in ab-initio calculations are relatively modest, the extra computation in the adopted procedure is negligible. 

\begin{algorithm}[H] \label{Algo:PseudochargeCalculation}
{\bf{Input}}: $\bR$, $V_J$, and $r_J^b$ \\
$b^{(i,j,k)}=0$, $E_{self}^{h,p}=0$\\
\For{$J \in P_{r_J^b}^p$}
{
Determine starting and ending indices $i_{s}$, $i_e$, $j_{s}$, $j_e$, $k_{s}$, $k_e$ for $\Omega_{r^b_J} \cap \Omega_p$ \\
Determine $V_J^{(i,j,k)}$ $\forall$ $i \in [i_s-n_o, i_e+n_o]$, $j \in [j_s - n_o, j_e+n_o]$, $k \in [k_s - n_o, k_e+n_o]$ \\
$b^{(i,j,k)}_J = - \frac{1}{4\pi} \nabla^2_h V_J \big|^{(i,j,k)}$, \quad $b^{(i,j,k)} = b^{(i,j,k)} + b^{(i,j,k)}_J $ $\forall$ $i \in [i_s,i_e]$, $j \in [j_s,j_e]$, $k \in [k_s,k_e]$ \\
$E_{self}^{h,p} = E_{self}^{h,p} + \frac{1}{2} h^3 b^{(i,j,k)}_J V_J^{(i,j,k)}$ $\forall$ $i \in [i_s,i_e]$, $j \in [j_s,j_e]$, $k \in [k_s,k_e]$ \\  
}
$E_{self}^h = \sum_{p=1}^{n_p} E_{self}^{h,p}$ \\ 
{\bf{Output}}: $b^{(i,j,k)}$ and $E_{self}^h$
\caption{Pseudocharge density generation and self energy calculation}
\end{algorithm} 


\subsection{Electrostatic potential calculation} \label{Subsec:PoissonEqn}
The electrostatic potential $\phi$---solution to the Poisson problem in Eqn. \ref{Eqn:EL} on all of space $\R^3$---needs to be computed in each SCF iteration as part of the linearized Hamiltonian $\mathbf{H}$ . However, since all calculations are restricted to $\Omega$, appropriate boundary conditions need to be prescribed on $\partial \Omega$ in order to minimize the finite-domain effect. Indeed, the simplest choice of zero Dirichlet boundary conditions can result in very slow convergence with domain size, as is evident from the discussion that follows. The electrostatic potential can be written in integral form using the Green's function of the Laplacian: 
\begin{equation} \label{Eqn:Phi:redsFunction}
\phi(\bx) = \int_{\R^3} \frac{\rho(\bx')+b(\bx',\bR)}{|\bx - \bx'|} \, {\rm d\bx'} \approx \int_{\Omega} \frac{\rho(\bx')+b(\bx',\bR)}{|\bx - \bx'|} \, \rm{d\bx'} \,, 
\end{equation}
where the exponential decay of the electron density $\rho$ and total pseudocharge $b$ has been used to restrict the integral to $\Omega$. On performing a multipole expansion of the kernel $1/|\bx-\bx'|$, we arrive at
\begin{equation} \label{Eqn:Phi:Multipole}
\phi(\bx) = \sum_{l=0}^{\infty} \sum_{m=-l}^{l} \frac{4\pi}{(2l+1)|\bx|^{l+1}}  Y_{lm}\left(\frac{\bx}{|\bx|}\right) \int_{\Omega} |\bx'|^l Y_{lm}\left(\frac{\bx'}{|\bx'|}\right) (\rho(\bx') + b(\bx')) \rm{d\bx'} \,\,,
\end{equation} 
where $Y_{lm}$ are the real spherical harmonics. It can therefore be deduced that unlike $\rho$ and $b$, in general $\phi$ only has algebraic decay away from the cluster. Therefore, significant errors can result when zero Dirichlet boundary conditions are employed, particularly for systems with net charge and/or dipole moment. In order to mitigate this, we adopt the procedure described below. 
We write the discrete form of the Poisson problem in Eqn. \ref{Eqn:EL} as\footnote{The `charge correction' is introduced into the Poisson equation to enforce the boundary conditions arising from Eqn. \ref{Eqn:Phi:redsFunction} within the high-order finite-difference method. Specifically, the `charge correction' takes non-zero values in a discontinuous fashion close to the boundary $\partial \Omega$, as can be inferred from Eqn. \ref{Eqn:Phi:correction}. This results in the solution of Eqn. \ref{Eqn:FD:PoissonDiscrete} subject to zero-Dirichlet boundary conditions being discontinuous at $\partial \Omega$. However, the electrostatic potential so calculated is continuous with respect to the correct boundary conditions arising from Eqn. \ref{Eqn:Phi:redsFunction}, and therefore possesses the desired accuracy. Overall, this technique implements the finite-domain boundary conditions on the electrostatic potential in the framework of high-order finite-differences.}
\begin{equation}\label{Eqn:FD:PoissonDiscrete}
-\frac{1}{4 \pi} \nabla_h^2 \phi \big|^{(i,j,k)} =  \rho^{(i,j,k)} + b^{(i,j,k)} - d^{(i,j,k)} \,,
\end{equation}
where zero Dirichlet boundary conditions are prescribed on $\partial \Omega$, and the `charge correction' \cite{burdick2003parallel} 
\begin{eqnarray}
d^{(i,j,k)} & = & \frac{-1}{4 \pi} \sum_{p=0}^{n_o} w_p \bigg(\chi^{(i+p,j,k)} \phi^{(i+p,j,k)} + \chi^{(i-p,j,k)} \phi^{(i-p,j,k)} + \chi^{(i,j+p,k)} \phi^{(i,j+p,k)} + \chi^{(i,j-p,k)} \phi^{(i,j-p,k)} \nonumber \\
& + & \chi^{(i,j,k+p)} \phi^{(i,j,k+p)} + \chi^{(i,j,k-p)} \phi^{(i,j,k-p)} \bigg) \label{Eqn:Phi:correction} \,. 
\end{eqnarray}
In the above expression, $w_p$ are the finite-difference weights given by Eqn. \ref{Eqn:weightsLap}, and $\chi$ is the indicator function that takes values of $0$ and $1$ when the index does and does not belong to the finite-difference grid, respectively. The values of $\phi^{(i,j,k)}$ corresponding to $\chi^{(i,j,k)}=1$ are calculated using the discrete truncated version of the multipole expansion in Eqn. \ref{Eqn:Phi:Multipole}:
\begin{equation} \label{Eqn:phi:BC:discrete}
\phi^{(i,j,k)} =  \sum_{l=0}^{l_{max}} \sum_{m=-l}^{l}  \frac{4\pi}{(2l+1)|\bx^{(i,j,k)}|^{l+1}} Y_{lm}^{(i,j,k)} Q_{lm}^h \,,
\end{equation}
where $l_{max}$ is the maximum angular momentum component, and the discrete multipole moments 
\begin{equation}
Q_{lm}^h =  h^3  \sum_{r=1}^{n_1} \sum_{s=1}^{n_2} \sum_{t=1}^{n_3} |\bx^{(r,s,t)}|^l Y_{lm}^{(r,s,t)} (\rho^{(r,s,t)} + b^{(r,s,t)}) \,.
\end{equation}

It is worth noting that the evaluation of $Q_{lm}$ is independent of the position at which the electrostatic potential needs to be evaluated. Therefore, the cost of calculating the charge correction is $\mathcal{O}(N_d) + \mathcal{O}(N_d^{2/3})$, which makes its scaling $\mathcal{O}(N)$ with respect to the number of atoms. The associated prefactors are insignificant since $l_{max}$ is typically very small, and $d^{(i,j,k)}$ only needs to be computed for grid points which lie within a distance of $(n_o-1)h$ from the boundary $\partial \Omega$. Therefore, the electrostatic potential $\phi$ can be determined in $\mathcal{O}(N)$ time when sophisticated preconditioners like multigrid \cite{hackbusch2013multi} are employed for solving the linear system in Eqn. \ref{Eqn:FD:PoissonDiscrete}. The above strategy is expected to minimize the finite-domain effect resulting from the slow decay of the electrostatic potential, which is indeed verified by the results presented in Section \ref{Sec:Examples}. 

\subsection{Electron density calculation} \label{Subsec:Rho}
In each iteration of the SCF method, the electron density corresponding to the linearized Hamiltonian $\mathbf{H}$ needs to be evaluated. This is typically the most computationally expensive step in DFT calculations. In this work, we utilize the Chebyshev filtered subspace iteration (CheFSI) \cite{zhou2006self,zhou2006parallel} to compute approximations to the lowest $N_s$ eigenvalues and corresponding eigenvectors of $\mathbf{H}$. This choice of eigensolver is motivated by the minimal orthogonalization and computer memory costs compared to other eigensolvers commonly employed in electronic structure calculations, e.g. Locally Optimal Block Preconditioned Conjugate Gradient (LOBPCG) \cite{knyazev2001toward}. Moreover, the lack of efficient real-space preconditioners limits the effectiveness of diagonalization approaches like LOBPCG in the current setting. In fact, CheFSI has been found to outperform LOBPCG for large scale computations even in the context of plane-waves \cite{levitt2015parallel}. 

The CheFSI algorithm as implemented in SPARC consists of three main steps. First, we filter the guess orbitals $\bf \Psi$ using Chebyshev polynomials:
\begin{equation} \label{Eqn:ChebFilter}
{\bf{\Psi}_f} = p_m({\bf{H}}){\bf \Psi} \,, \quad p_m(t) = C_m \left( \frac{t-c}{e} \right) \,,
\end{equation}
where ${\bf{\Psi}_f}$ represents the collection of filtered orbitals, and $C_m$ denotes the Chebyshev polynomial of degree $m$. In addition, $e = (\lambda_{N_d} - \lambda_c)/2$ and $c = (\lambda_{N_d} + \lambda_c)/2$, where $\lambda_c$ signifies the cutoff chosen for the Chebyshev polynomial filter. The central idea of this technique is to use the rapid growth of Chebyshev polynomials outside the interval $[-1,1]$ to dampen all the eigencomponents corresponding to eigenvalues larger than $\lambda_c$. The matrix $p_m({\bf{H}})$ is not explicitly determined, rather its product with $\mathbf{\Psi}$ is computed using the three term recurrence relation of Chebyshev polynomials, as outlined in Algorithm \ref{Algo:ChebychevFilter}.
\begin{algorithm}[h!] 
{\bf{Input}}: $\bf{H}$, ${\bf{\Psi}}$, $m$, $\lambda_1$, $\lambda_{N_d}$, $\lambda_c$  \\
$e = \frac{\lambda_{N_d} - \lambda_c}{2}$; $c = \frac{\lambda_{N_d} + \lambda_c}{2}$; $\sigma = \frac{e}{\lambda_1 - c}$ \\
 ${\bf{\Psi}_f} = \frac{\sigma}{e}\left({\bf{H}} - c {\bf{I}} \right ) {\bf{\Psi}}$ \\
\For{$j = 2:m$}
{
 ${\bf{\tilde{\Psi}}_f} = \frac{2\sigma}{e}\left({\bf{H}} - c {\bf{I}} \right) {\bf{\Psi}_f} - \left( \frac{\sigma^2}{2 - \sigma^2}\right) {\bf{\Psi}}$ \\
 ${\bf{\Psi}} = {\bf{\Psi}_f}$; \, ${\bf{\Psi}_f} = {\bf{\tilde{\Psi}}_f}$; \, $\sigma = \frac{\sigma}{2 - \sigma^2}$\\
}
{\bf{Output}}: ${\bf{\Psi}_f}$ 
\caption{Chebyshev filtering}
\label{Algo:ChebychevFilter}
\end{algorithm} 

Next, we project onto the filtered basis $\mathbf{\Psi_f}$ to arrive at the generalized eigenproblem: 
\begin{equation}\label{Eqn:SubspaceGeneralizedEigenproblem}
{\mathbf{H_s}} \mathbf{y_n} = \lambda_n \mathbf{M_s} \mathbf{y_n} \,, \quad  n=1,2, \ldots N_s \,,
\end{equation}
whose eigenvalues represent approximations to those of the Hamiltonian $\mathbf{H}$. The dense matrices $\mathbf{H_s}, \mathbf{M_s} \in \R^{N_s \times N_s} $ are obtained using the relations
\begin{equation}
\mathbf{H_s} = \mathbf{\Psi_f^T} \mathbf{H} \mathbf{\Psi_f} \,, \quad \mathbf{M_s} = \mathbf{\Psi_f^T} \mathbf{\Psi_f} \,.
\end{equation}
After solving the eigenproblem in Eqn. \ref{Eqn:SubspaceGeneralizedEigenproblem}, we calculate the Fermi energy $\lambda_f$ by enforcing the constraint on the total number of electrons:
\begin{equation} \label{Eqn:FermiEnergyConstraintFormulation}
2 \sum_{n=1}^{N_s} g_n = N_e \,, \quad \text{where} \quad g_n = \left(1 + \exp\left( \frac{\lambda_n - \lambda_f}{k_B T} \right) \right)^{-1} \,.
\end{equation}
Finally, we perform the subspace rotation 
\begin{equation}\label{Eqn:OrbitalTransformation}
\mathbf{\Psi} = \mathbf{\Psi_f} \mathbf{Y} \,, 
\end{equation}  
where the columns of the matrix $\mathbf{Y} \in \R^{N_s \times N_s}$ contain the eigenvectors $\mathbf{y_n}$. The columns of $\mathbf{\Psi}$ so obtained represent approximations to the eigenvectors of $\mathbf{H}$, which are then used to calculate the electron density at the finite-difference grid points: 
\begin{equation}
\rho^{(i,j,k)} =  \frac{2}{h^3}\displaystyle\sum_{n=1}^{N_s} \, g_n \, \psi_n^{2(i,j,k)} \,, 
\end{equation}
where $\psi_n^{(i,j,k)}$ are extracted from the $n^{th}$ column of $\mathbf{\Psi}$.

In the very first SCF iteration of the complete DFT simulation, we start with a randomly generated $\mathbf{\Psi}$, and repeat the steps in CheFSI---without calculating/updating the electron density---multiple times ($\sim 3$) \cite{zhou2014chebyshev}. This allows us to obtain a good approximation of the electron density for the second SCF iteration. In principle, this can be achieved by performing the CheFSI steps only once but with a higher degree Chebyshev polynomial $m$. However, in practice this causes the orbitals to become linearly dependent, which is prevented in the current procedure by the orthogonalization step within CheFSI. In every subsequent SCF iteration, we perform the CheFSI steps only once with the subspace rotated $\bf\Psi$ from the previous step as the initial guess. Overall, the calculation of the electron density scales as $\mathcal{O}(N_s N_d) + \mathcal{O}(N_s^2 N_d) + \mathcal{O}(N_s^3)$, which makes it $\mathcal{O}(N^3)$ with respect to the number of atoms. 

\subsection{Free energy calculation}
We approximate the integrals in Eqn. \ref{Eqn:FreeEnergy:GroundState:DFT} using the integration rule in Eqn. \ref{Eqn:IntApprox} to arrive at the following expression for the discrete free energy:
\begin{eqnarray} 
\mathcal{\hat{F}}^{h} & = & 2\sum_{n=1}^{N_s} g_n \lambda_n + h^3 \sum_{i=1}^{n_1} \sum_{j=1}^{n_2} \sum_{k=1}^{n_3} \bigg( \varepsilon_{xc}^{(i,j,k)}  \rho^{(i,j,k)} - V_{xc}^{(i,j,k)}\rho^{(i,j,k)} + \frac{1}{2} (b^{(i,j,k)}-\rho^{(i,j,k)}) \phi^{(i,j,k)} \bigg) \nonumber \\
& & - E_{self}^h + E_c^h + 2 k_B T \sum_{n=1}^{N_{s}} \left( g_n \log g_n + (1-g_n) \log (1-g_n) \right)   \label{Eqn:FreeEnergy:GroundState:DFT:Discrete} \,,
\end{eqnarray}
where $E_{self}^h$ is the discrete self energy of the pseudocharges (Eqn.~\ref{Eqn:DiscreteEself}), and $E_c^h$ is the discrete repulsive energy correction due to overlapping pseudocharges (Eqn.~\ref{Eqn:Ec:Disc}). The evaluation of $\mathcal{\hat{F}}^{h}$ scales as $\mathcal{O}(N_d)$, and therefore $\mathcal{O}(N)$ with respect to the number of atoms. Even though the free energy needs to be calculated only after the electronic/structural ground-state is determined, it is computed during each step of the SCF method, as is common practice in electronic structure calculations. 

\subsection{Atomic forces calculation}
The discrete form of the atomic force presented in Eqn. \ref{Eqn:Force:Nuclei} is the sum of three components:
\begin{equation}
\mathbf{f}_J^h = \mathbf{f}_{J,loc}^h + \mathbf{f}_{J,c}^h + \mathbf{f}_{J,nloc}^h \,,
\end{equation}
where $\mathbf{f}_{J,loc}^h$ is the discrete local component of the force, $\mathbf{f}_{J,c}^h$ is the discrete electrostatic correction for overlapping pseudocharges, and $\mathbf{f}_{J,nloc}^h$ is the discrete non-local component of the force. Below, we present expressions for $\mathbf{f}_{J,loc}^h$ and $\mathbf{f}_{J,nloc}^h$, and discuss their evaluation in SPARC. The expression for $\mathbf{f}_{J,c}^h$ can be found in Eqn. \ref{Eqn:LF:Disc}, and its evaluation progresses along similar lines as $\mathbf{f}_{J,loc}^h$. 

\paragraph{Local component} 
The local component of the atomic force in discrete form can be written as
\begin{equation} \label{Eqn:LocalForce:Discrete}
\mathbf{f}_{J,loc}^h = h^3  \sum_{i=1}^{n_1} \sum_{j=1}^{n_2} \sum_{k=1}^{n_3} \mathbf{\nabla}_h b_{J}\big|^{(i,j,k)} (\phi^{(i,j,k)} - V_{J}^{(i,j,k)}) \,,
\end{equation}
where the integral in Eqn. \ref{Eqn:Force:Nuclei} has been approximated using the integration rule in Eqn. \ref{Eqn:IntApprox}. The calculation of $\mathbf{f}_{J,loc}^h$ proceeds as outlined in Algorithm \ref{Algo:LocalForceCalculation}. Specifically, $V_J^{(i,j,k)}$ is interpolated on to the finite-difference grid in the overlap region $\Omega_{r_J^b} \cap \Omega_p \neq \emptyset$ (and an additional $4 n_0$ points in each direction) using cubic-splines, from which $b_J^{(i,j,k)}$ is calculated using Eqn. \ref{Eqn:PseudochargeDiscrete}. Subsequently, $\mathbf{f}_{J,loc}^{h,p}$---contribution of the $p^{th}$ processor to the local component of the force---is calculated using Eqn. \ref{Eqn:LocalForce:Discrete}. Finally, the contributions from all processors are summed to simultaneously obtain $\mathbf{f}_{J,loc}^h$ for all the atoms. 

\begin{algorithm}[H] \label{Algo:LocalForceCalculation}
\textbf{Input}: $\bR$, $\phi^{(i,j,k)}$, $V_J$, and $r_J^b$ \\
\For{$J \in P_{r_J^b}^p$}
{
Determine starting and ending indices $i_s$, $i_e$, $j_s$, $j_e$, $k_s$, $k_e$ for $\Omega_{r_J^b} \cap \Omega_p$ \\
Determine $V_J^{(i,j,k)}$ $\forall$ $i \in [i_s-2n_o, i_e+2n_o]$, $j \in [j_s - 2n_o, j_e+2n_o]$, $k \in [k_s - 2n_o, k_e+2n_o]$ \\
$b_J^{(i,j,k)} = - \frac{1}{4\pi} \nabla^2_h V_J\big|^{(i,j,k)}$ $\forall$ $i \in [i_s-n_o,i_e+n_o]$, $j \in [j_s-n_o,j_e+n_o]$, $k \in [k_s-n_o,k_e+n_o]$ \\  
$\mathbf{f}_{J,loc}^{h,p} = h^3 \sum_{i=i_s}^{i_e} \sum_{j=j_s}^{j_e} \sum_{k=k_s}^{k_e} \nabla_h b_J \big|^{(i,j,k)} ( \phi^{(i,j,k)} - V_J^{(i,j,k)} ) $ \\
}
$\mathbf{f}_{J,loc}^h = \sum_{p=1}^{n_p} \mathbf{f}_{J,loc}^{h,p}$ \\ 
\textbf{Output}: $\mathbf{f}_{J,loc}^h$
\caption{Calculation of the local component of the atomic force.}
\end{algorithm}

\paragraph{Non-local component}
The non-local component of the force in discrete form can be written as 
\begin{equation} \label{Eqn:NonlocForce:Discrete}
\mathbf{f}_{J,nloc}^h  =  -4\sum_{n=1}^{N_s} g_n \sum_{lm} \gamma_{Jl} Y_{Jnlm} \mathbf{W}_{Jnlm}   \,.
\end{equation}
where 
\begin{equation} \label{Eqn:NonLocForce:U:W}
Y_{Jnlm} = h^3  \sum_{i=1}^{n_1} \sum_{j=1}^{n_2} \sum_{k=1}^{n_3} \psi_n^{(i,j,k)} \chi_{Jlm}^{(i,j,k)} \,, \quad\mathbf{W}_{Jnlm} = h^3 \sum_{i=1}^{n_1} \sum_{j=1}^{n_2} \sum_{k=1}^{n_3} \nabla_h \psi_n \big|^{(i,j,k)}\chi_{Jlm}^{(i,j,k)} \,\,.
\end{equation}
Again, the integral in Eqn. \ref{Eqn:Force:Nuclei} has been approximated using the integration rule in Eqn. \ref{Eqn:IntApprox}. The calculation of $\mathbf{f}_{J,nloc}^h$ in SPARC is summarized in Algorithm \ref{Algo:non-localForceCalculation}. We use $P_{r_J^c}^p$ to denote the set of all atoms whose $\Omega_{r_J^c}$---cube with side of length $2r_J^c$ centered on the $J^{th}$ atom---has overlap with the processor domain $\Omega_p$. The value of $r_J^c$ corresponds to the maximum cutoff radius amongst the non-local components of the pseudopotential for the $J^{th}$ atom. We have chosen a cube rather than a sphere due to its simplicity and efficiency within the Euclidean finite-difference discretization. While describing Algorithm \ref{Algo:non-localForceCalculation}, we use the subscripts $s$ and $e$ to denote the starting and ending indices of $\Omega_{r_J^c} \cap \Omega_p \neq \emptyset$, respectively. In this overlap region, we interpolate the radial components of the projectors $\chi_{Jlm}^{(i,j,k)}$ on to the finite-difference grid using cubic-splines. Next, we utilize Eqn. \ref{Eqn:NonLocForce:U:W} to determine $Y_{Jnlm}^p$ and $\mathbf{W}_{Jnlm}^p$, which represent the contributions of the $p^{th}$ processor to $Y_{Jnlm}$ and $\mathbf{W}_{Jnlm}$, respectively. Finally, we sum the contributions from all the processors to obtain $Y_{Jnlm}$ and $\mathbf{W}_{Jnlm}$, which are then used to calculate $\mathbf{f}_{J,nloc}^h$ using Eqn. \ref{Eqn:NonlocForce:Discrete}. Overall, the calculation of the atomic forces scales as $\mathcal{O}(N)$ with respect to the number of atoms.

\begin{algorithm}[H] \label{Algo:non-localForceCalculation}
\textbf{Input}: $\bR$, $\psi_n^{(i,j,k)}$, $\gamma_{Jl}$, $\chi_{Jlm}$, and $r_J^c$ \\
$Y_{Jnlm}^p=0$, $\mathbf{W}_{Jnlm}^p=0$ \\
\For{$J \in P_{r_J^c}^p$}
{
 Determine starting and ending indices $i_s$, $i_e$, $j_s$, $j_e$, $k_s$, $k_e$ for $\Omega_{r_J^c} \cap \Omega_p$ \\
 Determine $\chi_{Jlm}^{(i,j,k)}$ $\forall$ $i \in [i_s,i_e]$, $j \in [j_s,j_e]$, $k \in [k_s,k_e]$ \\  
 $Y_{Jnlm}^p = Y_{Jnlm}^p + h^3 \psi_n^{(i,j,k)} \chi_{Jlm}^{(i,j,k)}$ $\forall$ $i \in [i_s,i_e]$, $j \in [j_s,j_e]$, $k \in [k_s,k_e]$ \\
 $\mathbf{W}_{Jnlm}^p = \mathbf{W}_{Jnlm}^p + h^3 \nabla_h \psi_n \big|^{(i,j,k)}\chi_{Jlm}^{(i,j,k)}$ $\forall$ $i \in [i_s,i_e]$, $j \in [j_s,j_e]$, $k \in [k_s,k_e]$ \\
}
$Y_{Jnlm} = \sum_{p=1}^{n_p} Y_{Jnlm}^p$, $\mathbf{W}_{Jnlm} = \sum_{p=1}^{n_p} \mathbf{W}_{Jnlm}^p$ \\ 
$\mathbf{f}_{J,nloc}^h = -4 \sum_{n=1}^{N_s} g_n \sum_{lm} \gamma_{Jl} Y_{Jnlm} \mathbf{W}_{Jnlm} $ \\
\textbf{Output}: $\mathbf{f}_{J,nloc}^h$
\caption{Calculation of the non-local component of the atomic force }
\end{algorithm}


\section{Examples and Results} \label{Sec:Examples}
In this section, we verify the proposed finite-difference formulation and parallel implementation of DFT for isolated clusters---first component of SPARC (Simulation Package for Ab-initio Real-space Calculations)---through selected examples.  In all the simulations, we utilize a twelfth-order accurate finite-difference discretization ($n_o=6$), the Perdew-Wang parametrization \cite{perdew1992accurate} of the correlation energy calculated by Ceperley-Alder \cite{Ceperley1980}, a smearing of $k_B T = 1 \times 10^{-3}$ Ha, and norm-conserving Troullier-Martins pseudopotentials \cite{Troullier}. The values of cutoff radii for the non-local projectors and the choice of local component of the pseudopotentials are specified in Appendix \ref{Appendix:Pseudopotential}. 

We truncate the discrete multipole expansion presented in Eqn. \ref{Eqn:phi:BC:discrete} at $l_{max}=6$. We solve the linear system in Eqn. \ref{Eqn:FD:PoissonDiscrete}---discrete form of the Poisson problem in Eqn. \ref{Eqn:EL}---using the Conjugate Gradient (CG) method \cite{cg1952} with the block-Jacobi preconditioner \cite{golub2012matrix}. In the CheFSI method, we determine the extremal eigenvalues of the Hamiltonian $\mathbf{H}$ using a few iterations of the Lanczos method \cite{lanczos1950iteration}, set the number of states to be $N_s = N_e/2 + 30$, utilize a polynomial of degree $m=20$ for Chebyshev filtering, and choose the filter cutoff $\lambda_c$ to be the previous iteration's Fermi energy plus $0.1$ Ha. Further, we solve the generalized eigenproblem in Eqn. \ref{Eqn:SubspaceGeneralizedEigenproblem} using the QR algorithm \cite{watkins2004fundamentals} as implemented in LAPACK \cite{laug}. We calculate the Fermi energy---root of the constraint in Eqn. \ref{Eqn:FermiEnergyConstraintFormulation}---using Brent's method \cite{press2007numerical}. We use Anderson mixing \cite{anderson1965iterative} with relaxation parameter of $0.3$ and mixing history of $7$ for accelerating the convergence of the Self-Consistent Field (SCF) method. Finally, we employ the Polak-Ribiere variant of non-linear conjugate gradients with a secant line search \cite{Shewchuk1994} for performing geometry optimization. 

In all the calculations, the energy and forces are converged to within the `chemical accuracy' of $0.001$ Ha/atom in energy and $0.001$ Ha/Bohr, respectively.\footnote{In this section, we will use the term `chemical accuracy' to denote the convergence of energy and atomic forces---with respect to the fully converged DFT results---to within $0.001$ Ha/atom and $0.001$ Ha/Bohr, respectively. It is worth emphasizing that DFT as a theory is an approximate one, and therefore in general does not necessarily produce chemically accurate results.} Wherever applicable, the results obtained by SPARC are compared to the well established plane-wave code ABINIT \cite{ABINIT,gonze2009abinit,gonze2005brief}. The error in energy is defined as the difference in the magnitude, and the error in forces is defined to be the maximum difference in any component on any atom. The simulations are performed on a  computer cluster consisting of $16$ nodes with the following configuration: Altus 1804i Server - 4P Interlagos Node, Quad AMD Opteron 6276, 16C, 2.3 GHz, 128GB, DDR3-1333 ECC, 80GB SSD, MLC, 2.5" HCA, Mellanox ConnectX 2, 1-port QSFP, QDR, memfree, CentOS, Version 5, and connected through InfiniBand cable.

\subsection{Convergence with domain size} 
\begin{figure}[H]
\centering
\subfloat[Energy]{\label{Fig:ConvergenceEnergyDomain} \includegraphics[keepaspectratio=true,width=0.46\textwidth]{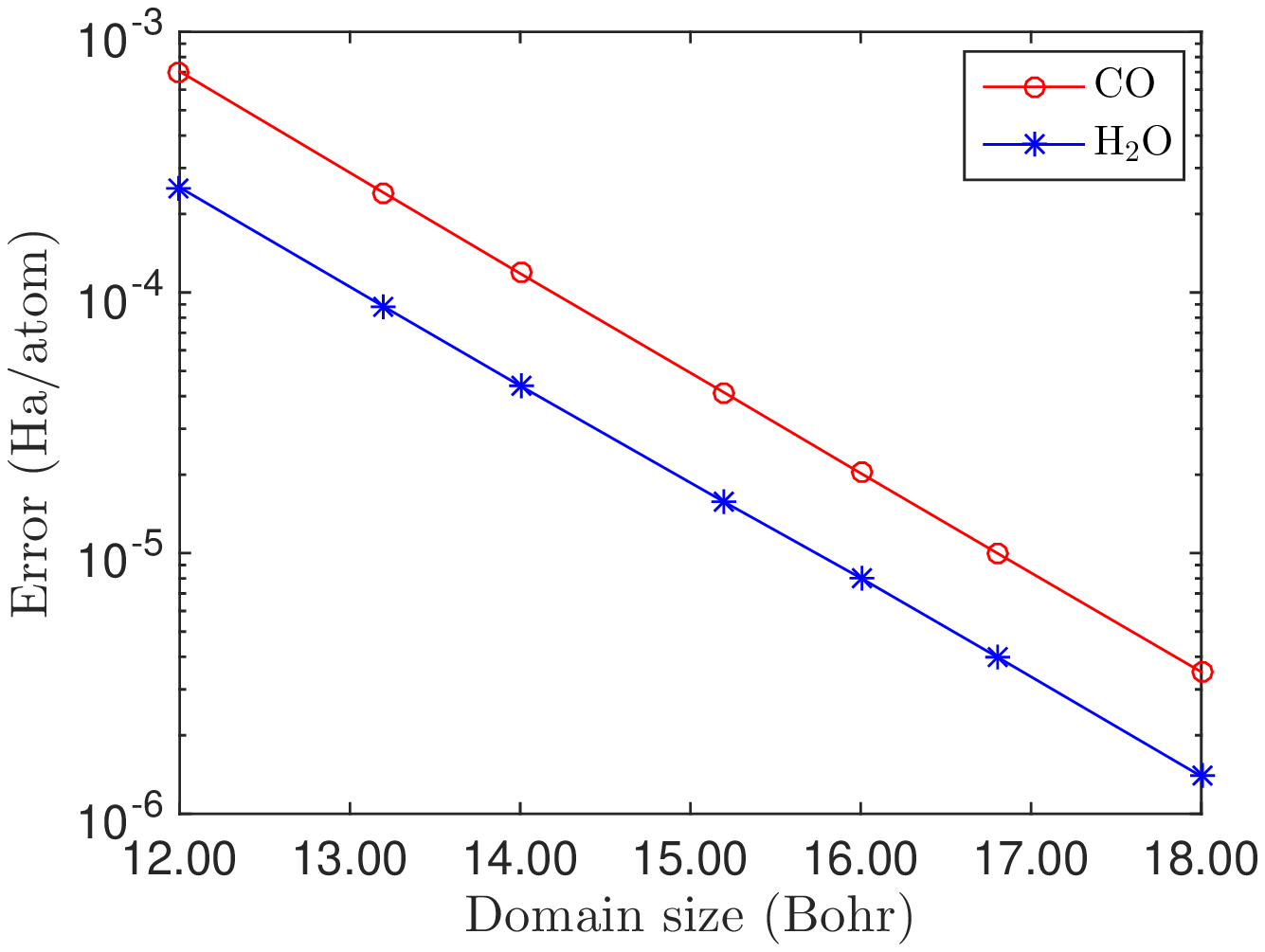}}
\subfloat[Forces]{\label{Fig:ConvergenceForceDomain} \includegraphics[keepaspectratio=true,width=0.46\textwidth]{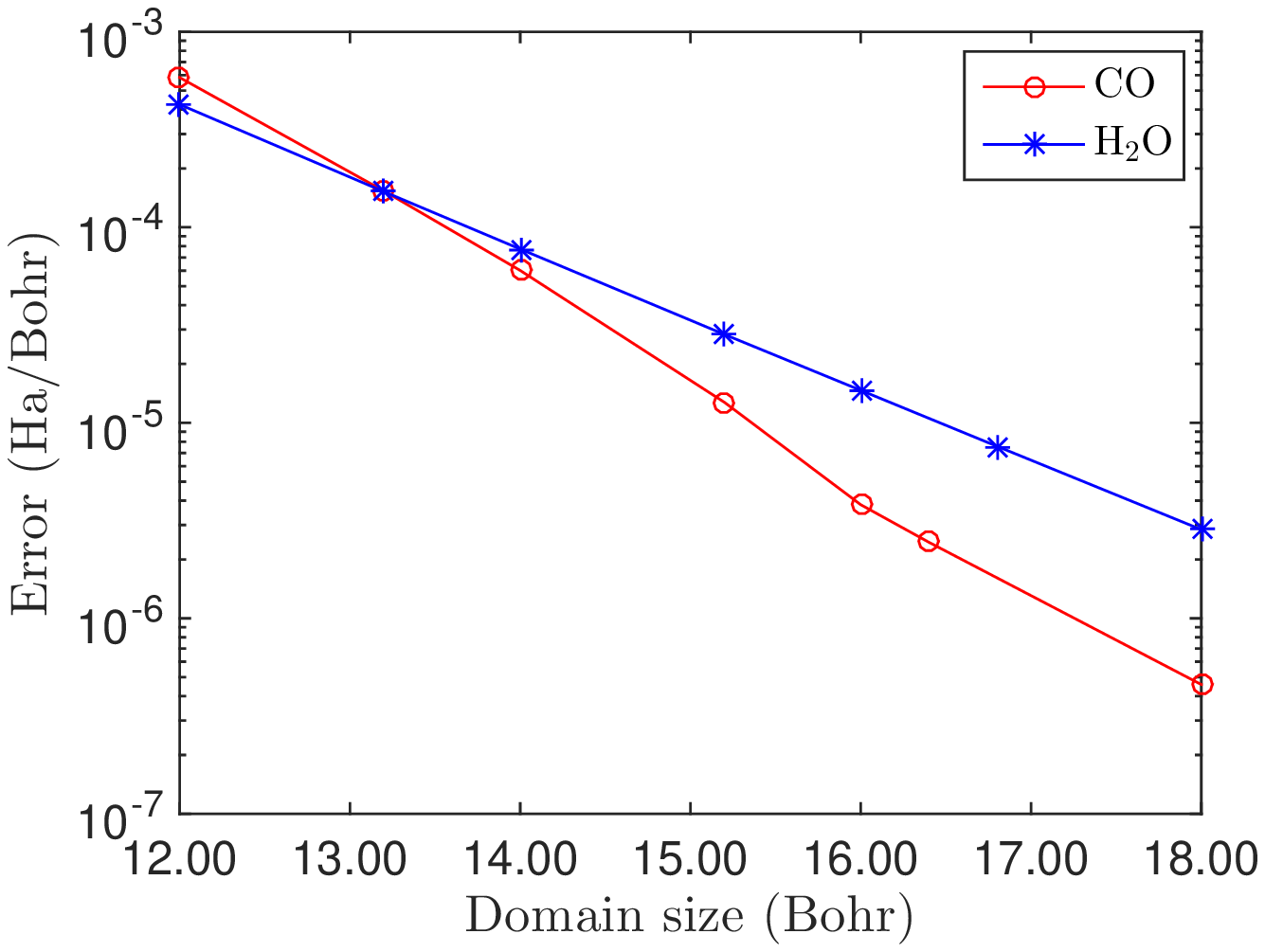}}
\caption{Convergence of energy and atomic forces with respect to domain size for the CO and H$_2$O molecules.}
\label{Fig:ConvergenceDomain}
\end{figure}

\begin{figure}[H]
\centering
\includegraphics[trim =0cm 1.6cm 0cm 0cm,clip,keepaspectratio=true,width=0.48\textwidth]{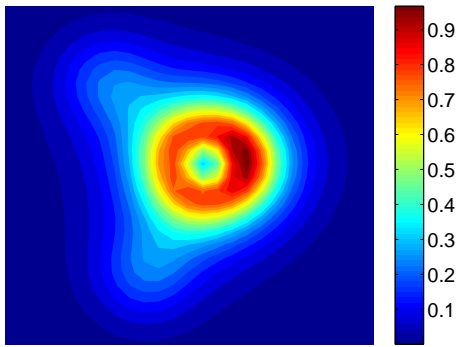}
\caption{In-plane electron density contours for the H$_2$O molecule.}
\label{Fig:WaterDensityContour} 
\end{figure}

We first verify the convergence of the computed energy and atomic forces with respect to the size of the domain $\Omega$. We choose the carbon monoxide (CO) and water (H$_2$O) molecules as representative examples, with the C-O and O-H bond lengths reduced and increased by $8 \%$ from their equilibrium values as determined by ABINIT, respectively. The polar nature of the molecules and their deliberate asymmetric positioning within $\Omega$ ensure that any finite-domain effects are exaggerated. In Fig. \ref{Fig:ConvergenceDomain}, we present convergence of the energy and atomic forces for $h=0.2$ Bohr as $\{L_1,L_2,L_3 \}$ is increased from $\{12,12,12 \}$ Bohr to $\{18,18,18\}$ Bohr, with the results obtained for $\{L_1,L_2,L_3 \} = \{40,40,40 \}$ Bohr used as reference. We observe exponential convergence of both the energy and the forces to well below accuracies desired in DFT calculations. In fact, even a domain size of $\{L_1,L_2,L_3 \} = \{12,12,12 \}$ is sufficient to obtain chemical accuracy in both energy and forces. The corresponding electron density contours for H$_2$O are plotted in Fig. \ref{Fig:WaterDensityContour}. Overall, these results demonstrate the efficacy of SPARC's electrostatic formulation in minimizing the finite-domain effect for isolated clusters. 

\subsection{Convergence with spatial discretization} \label{Subsec:ConvergenceMesh}
\begin{figure}[H]
\centering
\subfloat[Energy]{\label{fig:energyConvergence}\includegraphics[keepaspectratio=true,width=0.46\textwidth]{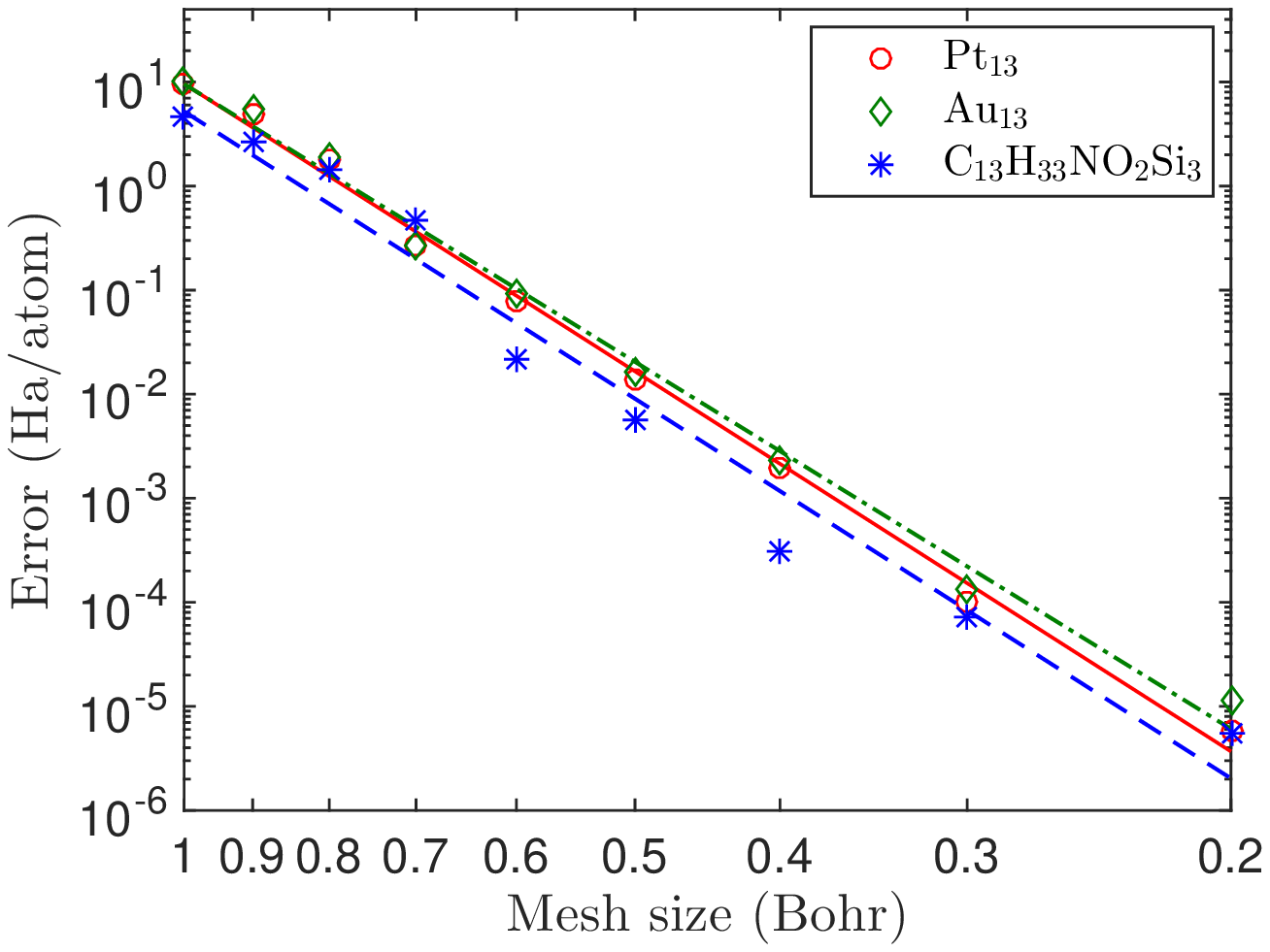}}
\subfloat[Forces]{\label{fig:forceConvergence}\includegraphics[keepaspectratio=true,width=0.46\textwidth]{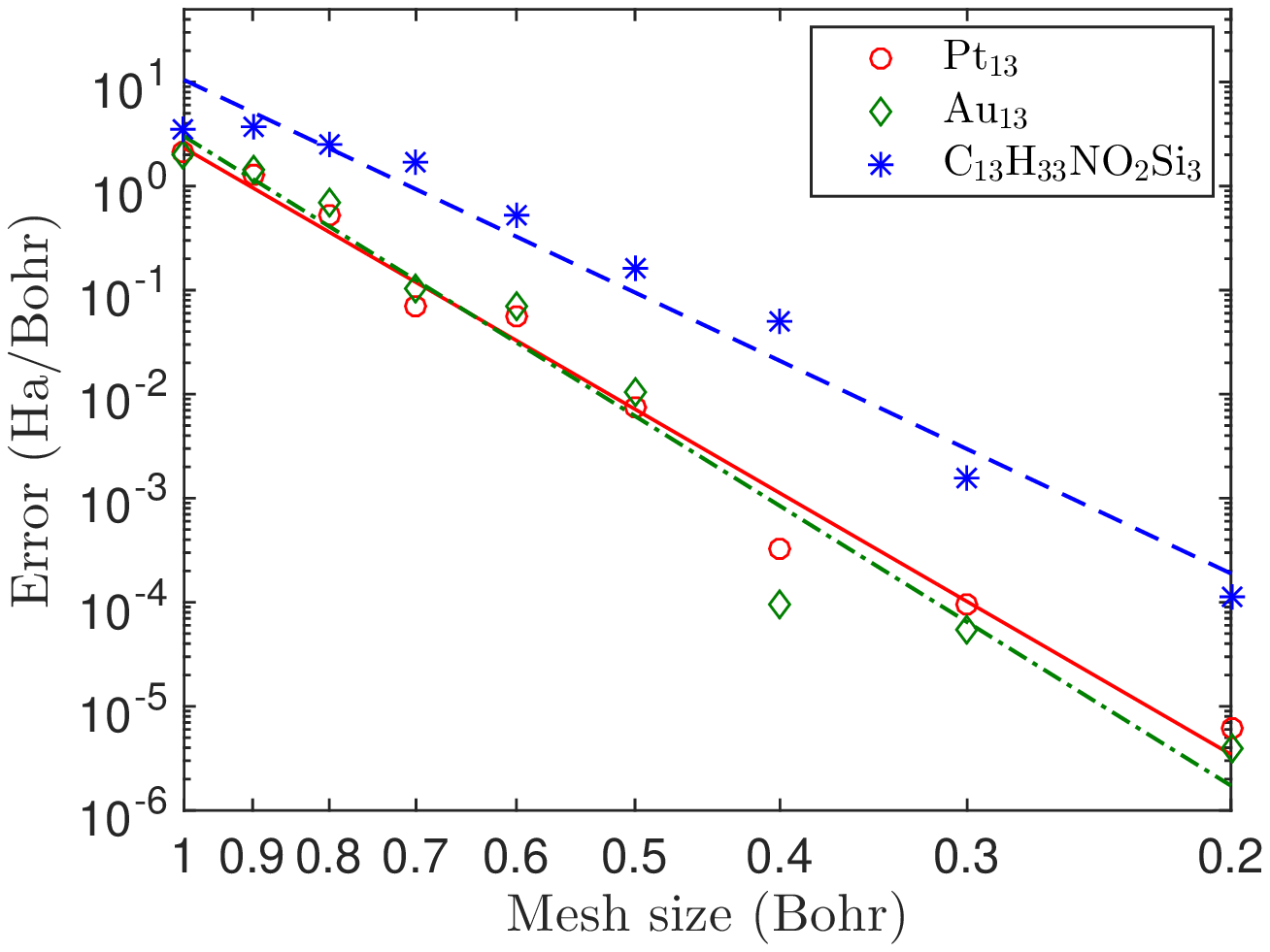}}
\caption{Convergence of the energy and atomic forces with respect to mesh size to reference planewave result for the Pt$_{13}$, Au$_{13}$, and C$_{13}$H$_{33}$NO$_2$Si$_3$ clusters.}
\label{fig:convergenceDiscretization}
\end{figure}

\begin{figure}[H]
\centering
\subfloat[Au$_{13}$]{\label{Fig:Au13Density}\includegraphics[trim =2.5cm 2.5cm 2.5cm 3.0cm, clip, keepaspectratio=true,width=0.36\textwidth]{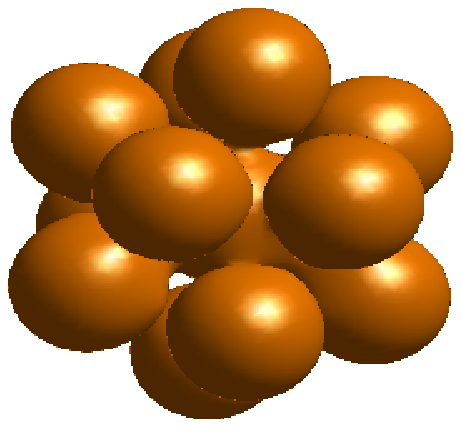}}
\hspace{5mm}
\subfloat[C$_{13}$H$_{33}$NO$_2$Si$_3$]{\label{Fig:BiomoleculeDensity}\includegraphics[trim =1.5cm 1cm 1cm 1.7cm, clip,keepaspectratio=true,width=0.45\textwidth]{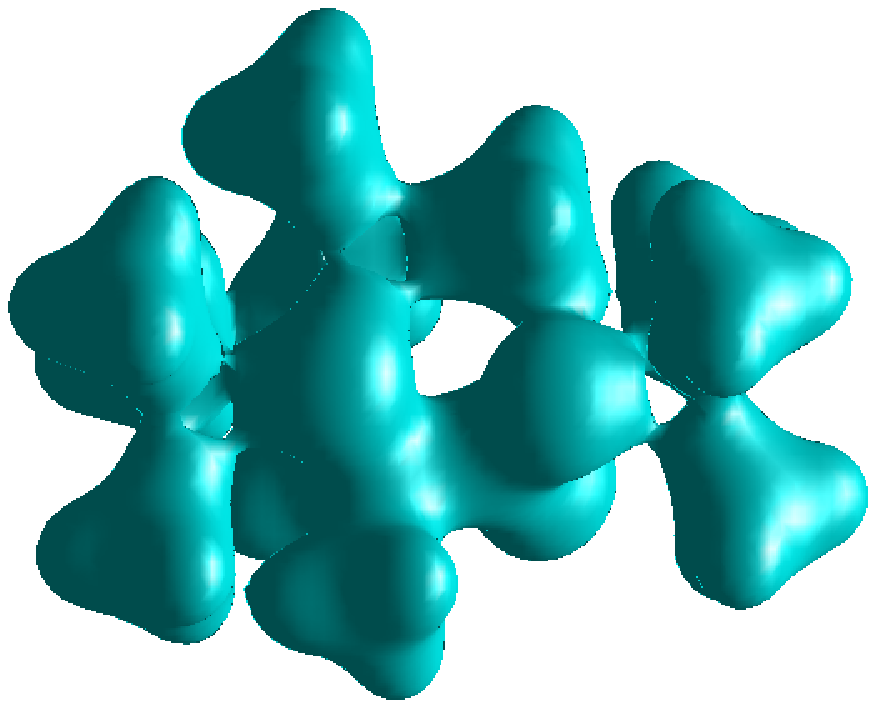}}
\caption{Electron density isosurface for $\rho=0.05$ Bohr$^{-3}$}
\label{Fig:Isosurface}
\end{figure}

We next verify the convergence of the computed energy and atomic forces with respect to the finite-difference mesh-size. As representative examples, we choose the icosahedral Platinum (Pt$_{13}$), icosahedral Gold (Au$_{13}$), and $\beta$-Aminoisobutyric acid tri-TMS II (C$_{13}$H$_{33}$NO$_2$Si$_3$) clusters with domain sizes of $\{L_1,L_2,L_3 \} = \{36,36,36\}$, $\{40,40,40\}$ and $\{42,38,34\}$ Bohr, respectively. All errors are defined with respect to ABINIT, wherein we employ plane-wave cutoffs of $42$, $42$, and $68$ Ha along with domain sizes of $\{L_1,L_2,L_3 \} =\{42,42,42\}$, $\{42,42,42\}$, and $\{50,46,42\}$ Bohr for Pt$_{13}$, Au$_{13}$, and C$_{13}$H$_{33}$NO$_2$Si$_3$, respectively. The resulting reference energies and forces are converged to within $5.0\times 10^{-6}$ Ha/atom and $5.0\times 10^{-6}$ Ha/Bohr, respectively. In Fig. \ref{fig:convergenceDiscretization}, we plot error in the SPARC energy and forces with respect to the mesh size, from which it is clear that there is systematic convergence of both energies and forces. On performing a fit to the data, we obtain average convergence rates of approximately $\mathcal{O}(h^{9})$ in the energy and $~\mathcal{O}(h^{8})$ in the forces. In doing so, the desired accuracy is readily attained. In Fig. \ref{Fig:Isosurface}, we present the computed isosurfaces for Au$_{13}$ and C$_{13}$H$_{33}$NO$_2$Si$_3$. Overall, we conclude that SPARC is able to obtain high convergence rates in both the DFT energy and atomic forces, which contributes to its accuracy and efficiency. Moreover, the energy and forces converge at comparable rates, without need of additional measures such as double-grid \cite{OnoHir99} or high-order integration \cite{BobSchChe15} techniques. 

\subsection{Ground state properties} \label{Subsec:GeometryOptimization}
\begin{figure}[H]
\centering
\subfloat[Computed energy and its cubic spline fit]{\label{Fig:energy_eggbox}\includegraphics[keepaspectratio=true,width=0.46\textwidth]{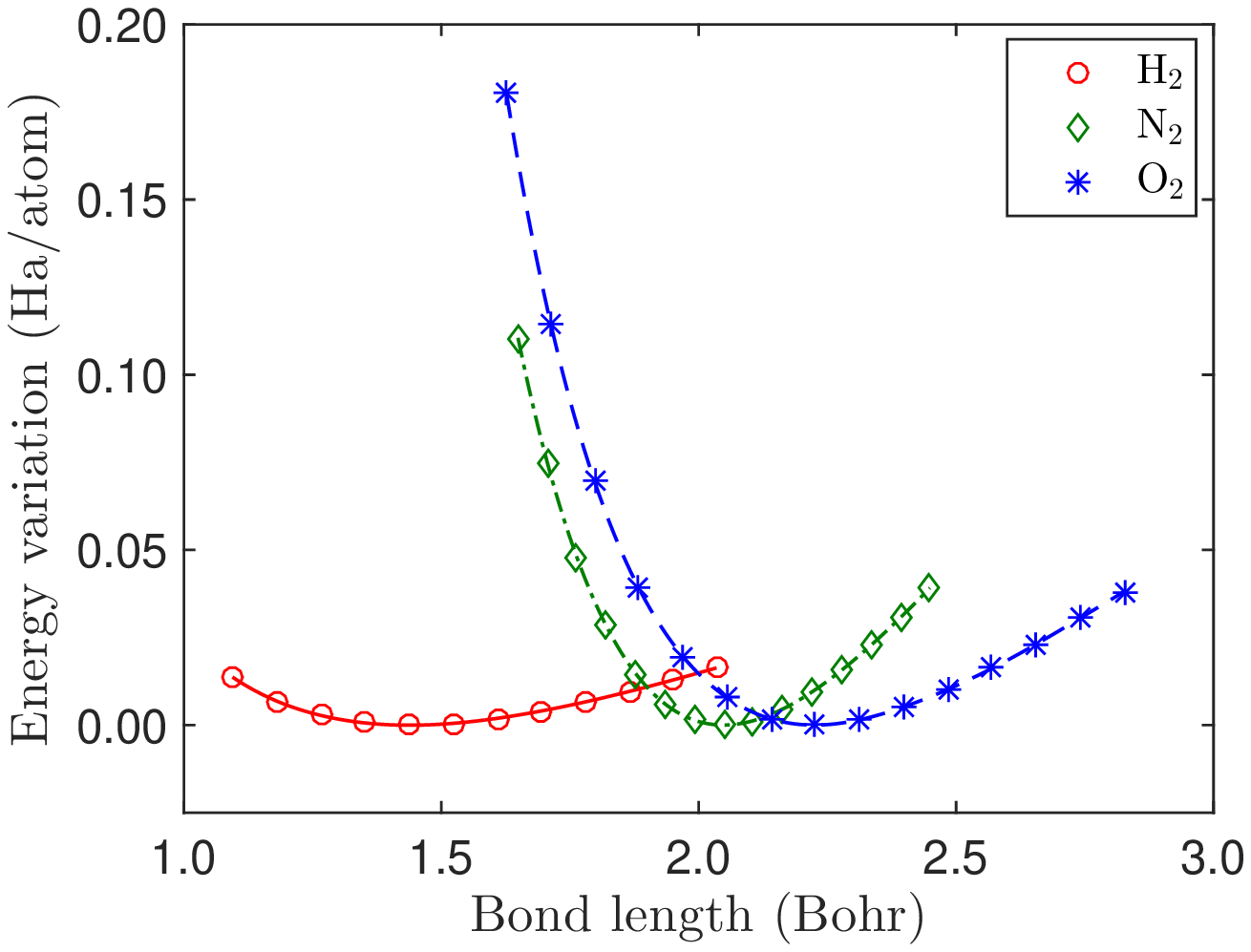}}
\subfloat[Computed force and the derivative of the cubic spline fit to the energy]{\label{Fig:energy_force_consistency}\includegraphics[keepaspectratio=true,width=0.46\textwidth]{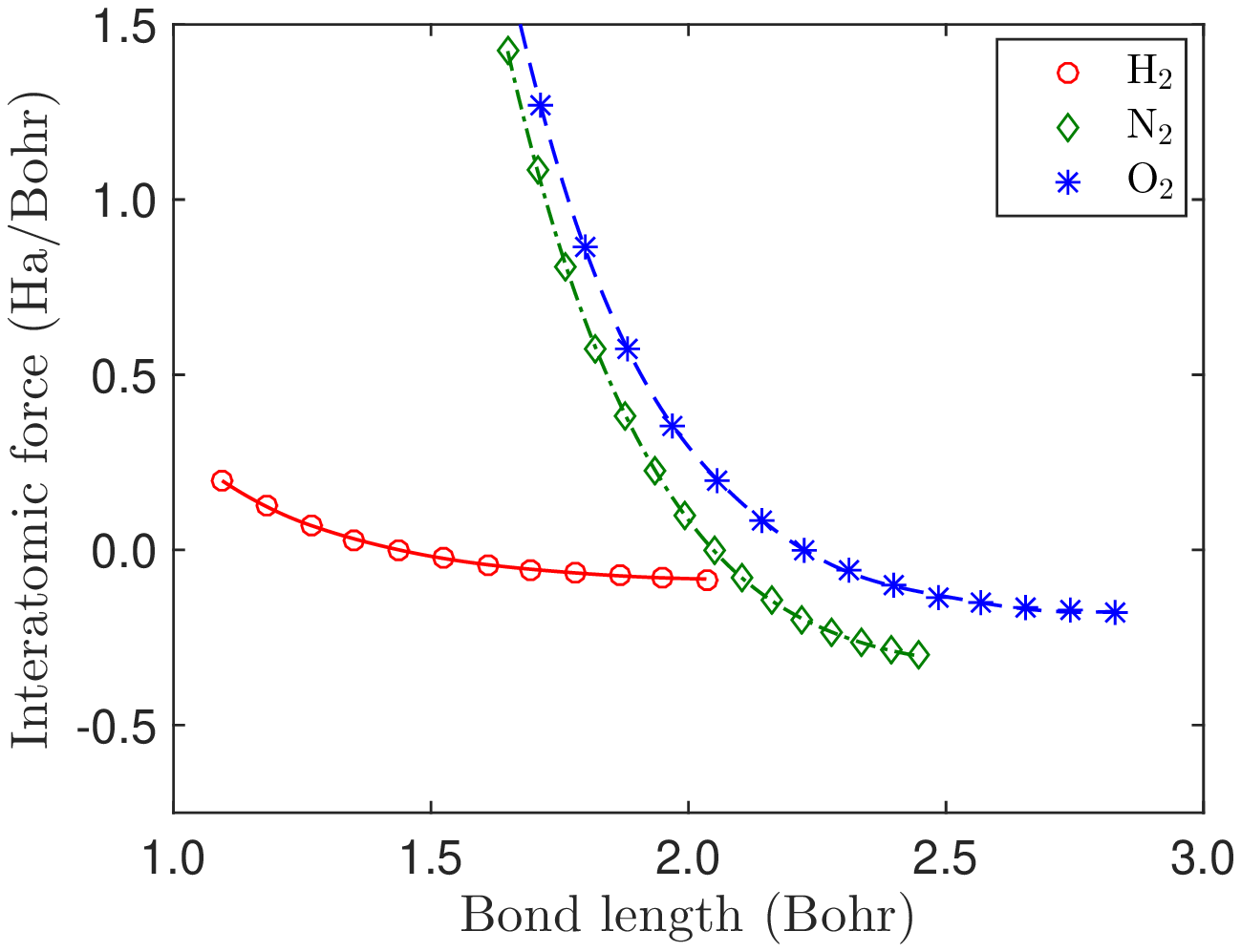}}
\caption{Variation in the computed energy and atomic force as a function of interatomic distance for the H$_2$, N$_2$, and O$_2$ molecules.}
\label{Fig:Eggbox:Consistency}
\end{figure}

We now verify that the ground-state properties of isolated clusters can be accurately determined using SPARC. For this purpose, we select the Hydrogen (H$_2$), Nitrogen (N$_2$), and Oxygen (O$_2$) molecules, a domain size of $\{L_1,L_2,L_3 \} = \{24,24,24\}$ Bohr, and mesh-size of $h=0.2$ Bohr. We begin by evaluating the energy and force as a function of interatomic distance, the results of which are presented in Fig. \ref{Fig:Eggbox:Consistency}. Specifically, we plot the energy as a function of bond length along with its cubic spline fit in Fig. \ref{Fig:energy_eggbox}, and the computed interatomic force and the derivative of the cubic spline fit to the energy in Fig. \ref{Fig:energy_force_consistency}. The evident agreement demonstrates that the computed energy and atomic forces are indeed consistent. Moreover, there is no noticeable `egg-box' effect \cite{brazdova2013atomistic}---a phenomenon arising due to the breaking of the translational symmetry---at meshes required for obtaining chemical accuracies.  

Next, we use the above results to calculate the vibrational frequency for the H$_2$, N$_2$, and O$_2$ molecules using the relation \cite{mohan2004organic}: 
\begin{equation}
\nu = \frac{1}{2\pi c} \left( \frac{k}{\mu} \right )^{\frac{1}{2}} \,,
\end{equation}
where $c$ is the speed of light, $k$ is the derivative of the cubic spline fit to the force at the equilibrium bond length, and $\mu$ is the reduced mass of the system. In ABINIT, we choose a domain size of $\{L_1,L_2,L_3 \} = \{30,30,30 \}$ Bohr for all three systems and planewave cutoffs of $32$ Ha, $40$ Ha, and $38$ Ha for H$_2$, N$_2$, and O$_2$, respectively. From the results presented in Table \ref{Table:Vibrational}, we observe that there is excellent agreement between SPARC and ABINIT, with the maximum difference in the vibrational frequency being $8$ $cm^{-1}$. There is also good agreement between DFT and experiment, highlighting the accuracy of DFT as an ab-initio theory. These results further verify that SPARC is able to obtain accurate atomic forces, a critical feature for both structural relaxations and ab-initio molecular dynamics. 

\begin{table}[H]
\centering
\begin{tabular}{cccc}
\hline 
Molecule & SPARC  & ABINIT &  Experiment \cite{huber2013,Becke1992}\\  
\hline
H$_2$        &  $4007$   & $4014$    &  $4401$               \\
N$_2$       &  $2448$    & $2456$    &  $2358$           \\
O$_2$       &  $1649$    & $1642$    &  $1580$               \\

\hline
\end{tabular}
\caption{Vibrational frequency in $cm^{-1}$ for the H$_2$, N$_2$, and O$_2$ molecules.}
\label{Table:Vibrational}
\end{table} 

Finally, we randomly perturb the atomic positions in the three molecules such that the interatomic distance differs by up to $15$ percent from the equilibrium bond length. We maximize generality by ensuring that the resulting systems are not aligned with any of the coordinate axes. In Table \ref{Table:GroundState}, we present the results of the geometry optimization by SPARC, and compare them with ABINIT for the aforementioned choice of parameters. We observe that there is very good agreement between SPARC and ABINIT, with the maximum difference in the energy being $0.0007$ Ha/atom, and the maximum difference in the equilibrium bond length being $0.001$ Bohr. These results are also in excellent agreement with the data plotted in Fig. \ref{Fig:Eggbox:Consistency}. Overall the results indicate that SPARC is able to accurately determine ground state properties for isolated clusters.

\begin{table}[H]
\centering
\begin{tabular}{ccccc}
\hline 
\multirow{ 2}{*}{Molecule} & \multicolumn{2}{c}{Energy (Ha/atom)} &  \multicolumn{2}{c}{Bond length (Bohr)} \\
         &  SPARC    & ABINIT   &   SPARC    & ABINIT \\
\hline
H$_2$        &  $-0.5682$   & $-0.5681$    &  $1.437$     &  $1.437$        \\
N$_2$       &  $-9.9463$    & $-9.9460$    &  $2.049$     & $2.049$   \\
O$_2$       &  $-15.8717$    & $-15.8710$    &  $2.226$    & $2.227$       \\

\hline
\end{tabular}
\caption{Ground state energy and equilibrium configuration for the H$_2$, N$_2$, and O$_2$ molecules.}
\label{Table:GroundState}
\end{table} 


\subsection{Scaling and performance} \label{Subsec:Scaling}
In previous subsections, we have verified the accuracy of SPARC by comparing with the well-established plane-wave code ABINIT. We now investigate the efficiency of SPARC relative to ABINIT, for which we choose bulk-terminated Silicon nanoclusters passivated by Hydrogen as representative examples. In all the calculations, we utilize a mesh-size of $h=0.5$ Bohr in SPARC, and a planewave energy cutoff of $16$ Ha in ABINIT. Further, we employ a vacuum of $5$ Bohr in both SPARC and ABINIT. We choose all the other parameters so as to obtain the chemical accuracy of $0.001$ Ha/atom in the energy and $0.001$ Ha/Bohr in the atomic force. All the times reported here include the calculation of the electronic ground-state as well as the atomic force. The detailed breakdown of the timings for SPARC can be found in the output files provided with the code accompanying this paper.

\begin{figure}[H]
\centering
\subfloat[Strong scaling]{\label{Fig:StrongScaling}\includegraphics[keepaspectratio=true,width=0.48\textwidth]{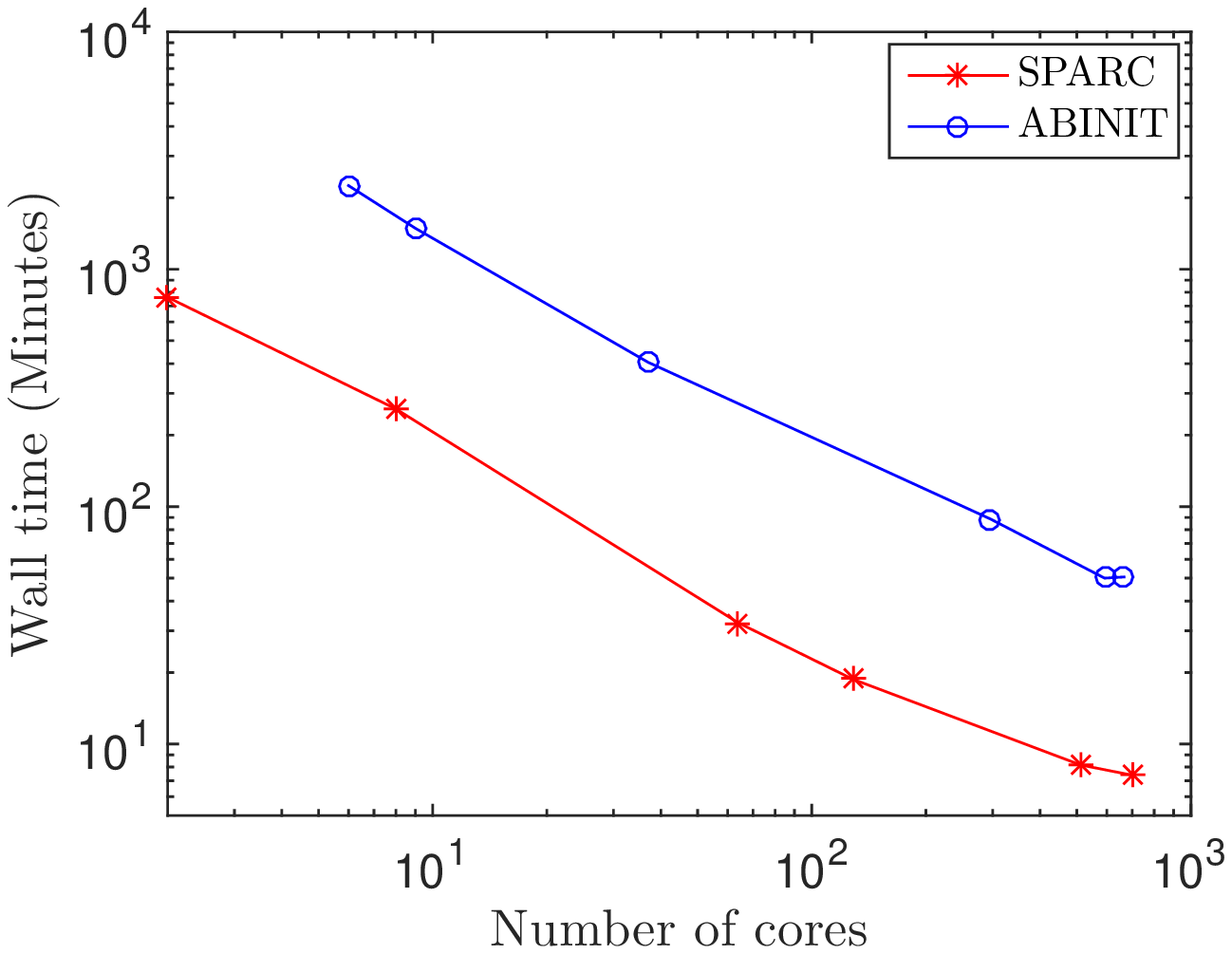}}
\subfloat[Weak scaling]{\label{Fig:WeakScaling}\includegraphics[keepaspectratio=true,width=0.48\textwidth]{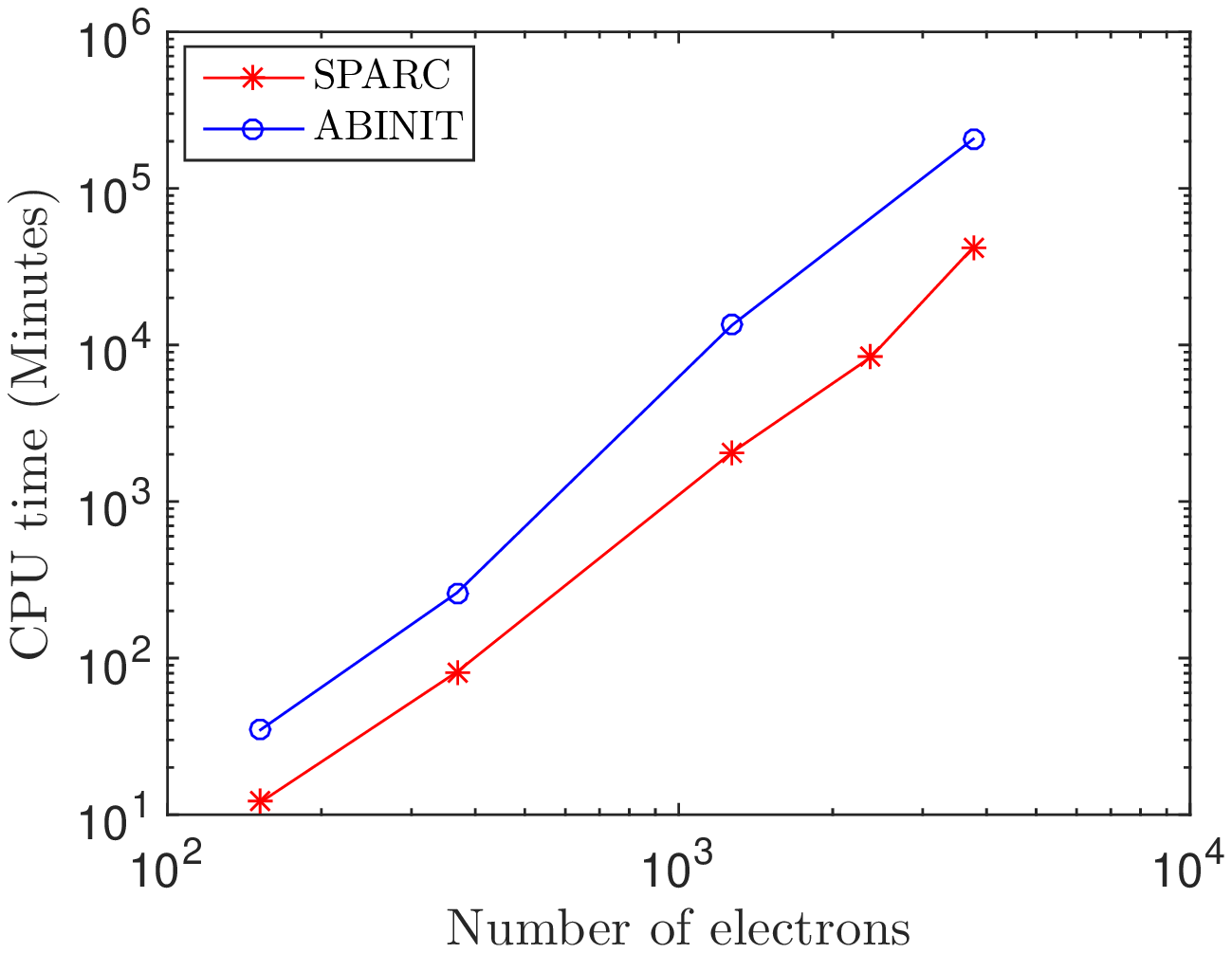}}
\caption{Strong and weak scaling behavior for hydrogen passivated silicon nanoclusters. The system utilized for strong scaling is Si$_{275}$H$_{172}$. The systems employed for weak scaling are Si$_{29}$H$_{36}$, Si$_{71}$H$_{84}$, Si$_{275}$H$_{172}$, Si$_{525}$H$_{276}$ and Si$_{849}$H$_{372}$.}
\end{figure}

First, we compare the strong scaling of SPARC with ABINIT for the Si$_{275}$H$_{172}$ cluster. We utilize $2$, $8$, $64$, $128$, $512$, and $640$ cores for performing the simulation using SPARC. We use  $6$, $9$, $37$, $296$, $592$, and $666$ cores for ABINIT, which it suggests are optimal in the range of cores considered here. Both SPARC and ABINIT require $19$ iterations for convergence of the SCF method. In Fig. \ref{Fig:StrongScaling}, we present the wall time taken by SPARC and ABINIT as the number of processors is increased. We observe that both SPARC and ABINIT display similar trends with respect to strong scaling. Specifically, the SPARC and ABINIT curves are close to being parallel, with no further reduction in wall time observed after approximately $700$ cores for SPARC and $600$ cores for ABINIT. However, the prefactors are significantly different, with SPARC being able to outperform ABINIT by up to factors of $7$.

Next, we compare the weak scaling of SPARC with ABINIT for the Si$_{29}$H$_{36}$, Si$_{71}$H$_{84}$, Si$_{275}$H$_{172}$, Si$_{525}$H$_{276}$, and Si$_{849}$H$_{372}$ nanoclusters. The number of electrons in these systems range from $152$ (Si$_{29}$H$_{36}$) to $3768$ (Si$_{849}$H$_{372}$). For both SPARC and ABINIT, we fix the number of electrons per core to be approximately $160$, and select at most $4$ cores from every compute node. In Figure \ref{Fig:WeakScaling}, we present the results so obtained for the variation in total CPU time versus the number of electrons.\footnote{SPARC's wall times for the Si$_{29}$H$_{36}$, Si$_{71}$H$_{84}$, Si$_{275}$H$_{172}$, Si$_{525}$H$_{276}$, and Si$_{849}$H$_{372}$ systems are $12$, $40$, $258$, $522$, and $1751$ mins, respectively. The corresponding wall times for ABINIT are $35$, $131$, $1489$, $3574$, and $8613$ mins, respectively.} We observe similar scaling for both codes, with $\mathcal{O}(N_e^{2.51})$ for SPARC and $\mathcal{O}(N_e^{2.75})$ for ABINIT. However, the prefactor for SPARC is again noticeably lower, with speedups over ABINIT ranging from factors of $3$ to $7$. 

The superior performance of SPARC relative to ABINIT for the examples considered here merits further consideration. In SPARC, more than $84 \%$ of the total time is spent in the Chebyshev filtering, subspace projection, and subspace rotation steps.\footnote{For the Si$_{29}$H$_{36}$ system, the Chebyshev filtering, projection, and subspace rotation steps take $77 \%$, $5.8 \%$ and $1.2 \%$ of the total time, respectively. For the Si$_{849}$H$_{372}$ system, the Chebyshev filtering, projection, and subspace rotation steps take $70 \%$, $19.1 \%$ and $8.8 \%$ of the total time, respectively.} In ABINIT, more than $90 \%$ of the time is spent in the subroutines \texttt{fourwf} (FFTs for the wavefunctions), \texttt{nonlop} (non-local pseudopotential related computations),\footnote{We have found that \texttt{nonlop} scales as $\sim \mathcal{O}(N^{2.6})$, and becomes the dominant cost for the larger systems considered here (e.g., \texttt{nonlop} takes $63 \%$ of the total time for Si$_{275}$H$_{172}$ in the weak scaling study). The closer to $\mathcal{O}(N^3)$ scaling suggests that ABINIT performs the nonlocal projections in reciprocal space, which can be made $\mathcal{O}(N^2)$ when implemented in real-space \cite{king1991real}, though at the cost of a modified pseudopotential and an increased prefactor. Even when the time taken for the nonlocal operations is excluded, SPARC is still able to outperform ABINIT, e.g., the speedup for the Si$_{29}$H$_{36}$ and Si$_{275}$H$_{172}$ clusters in the weak scaling study is $4.2$ and $1.8$, respectively.}  and \texttt{projbd} (Gram-Schmidt orthogonalizations). Since the number of orthogonalizations in the CheFSI method is very small---particularly when compared to LOBPCG---it is not a significant component in SPARC. Inspite of this, though the implementation of CheFSI in ABINIT is competitive compared to LOBPCG for some systems, it is slower for other systems, due to the lack of preconditioning in CheFSI\footnote{This is particularly the case for simulations on modest number of computational cores, as those employed in this work.} \cite{levitt2015parallel}. Therefore, it appears that the main reason for the speedup of SPARC over ABINIT is that the Hamiltonian operator can be applied more efficiently within the finite-difference discretization.\footnote{This is also expected to be the case for other real-space discretizations with highly localized orthonormal basis functions (i.e., identity overlap matrix), e.g., spectral finite-elements \cite{batcho2000computational,motamarri2012higher}. Even so, the finite-difference method is particularly attractive because of a number of reasons. First, the cost of performing integrations within the finite-difference method is practically zero. Second, for a given order of convergence, the finite-difference representation of the Laplacian is very compact. Finally, the spectral widths of the resulting Hamiltonians are relatively small compared to the alternatives. These features make the finite-difference approach an efficient choice for performing DFT calculations, as demonstrated by the examples in this work. }

Finally, we compare the minimum wall time achievable by SPARC and ABINIT for the aforementioned nanoclusters with the exception of Si$_{849}$H$_{372}$, for which the resources currently available to us are insufficient. While performing this study, we restrict the maximum number of electrons per computational core to $160$. In SPARC, we choose the number of cores as multiples of $64$, whereas we select the number of cores and parallelization scheme in ABINIT as suggested by it. We present the results so obtained in Table \ref{Table:Time}. We observe that SPARC is able to achieve smaller wall times by factors larger than $6.8$ compared to ABINIT for all the systems considered. In particular, SPARC requires a factor of approximately $7.0$ less wall time than ABINIT for the Si$_{525}$H$_{276}$ nanocluster. Overall, these results indicate that SPARC is a highly efficient DFT formulation and implementation that is highly competitive with well-optimized plane-wave codes.\footnote{This is also found to be the case when SPARC is compared to well-established finite-difference codes, as demonstrated in Appendix \ref{Appendix:RealSpaceComparison}.}

\begin{table}[H]
\centering
\begin{tabular}{ccc}
\hline
System &   SPARC     & ABINIT  \\   
\hline
Si$_{29}$H$_{36}$        &   $0.69$ $(128)$  & $7.30$ $(106)$     \\
Si$_{71}$H$_{84}$        &   $1.15$ $(320)$  & $8.10$ $(321)$    \\
Si$_{275}$H$_{172}$      &   $7.39$ $(704)$  & $50.60$ $(666)$    \\
Si$_{525}$H$_{276}$      &   $32.45$ $(960)$ & $227.56$ $(1008)$ \\
\hline
\end{tabular}
\caption{Minimum wall time in minutes for hydrogen passivated silicon nanoclusters. The number in brackets represents the number of cores on which the minimum wall time is achieved.}
\label{Table:Time}
\end{table}

\section{Concluding Remarks} \label{Sec:Conclusions} 
In this work, we have developed an accurate and efficient finite-difference formulation and parallel implementation of Density Functional Theory (DFT) for isolated clusters, which represents the first component of SPARC (Simulation Package for Ab-initio Real-space Calculations). Specifically, employing the Chebyshev polynomial filtered self-consistent field iteration in conjunction with the reformulation of the electrostatics and the non-local component of the atomic force, we have developed a framework using the finite-difference representation wherein energies and forces can be efficiently evaluated to within the accuracies desired in electronic structure calculations. Through a variety of examples consisting of both light and heavy elements, we have demonstrated that SPARC obtains exponential convergence in energies and forces with domain size; systematic convergence in the energy and forces with respect to spatial discretization at comparably high rates to reference plane-wave results; forces that are consistent with the energy, both being free from any noticeable `egg-box' effect; and accurate ground-state properties like equilibrium energies, geometries and vibrational spectra. Moreover, we have shown that the weak and strong parallel scaling of SPARC is very similar to well-established and optimized plane-wave codes for systems consisting of up to thousands of electrons, but with a significantly smaller prefactor. 

The examples in this work have been restricted to $\sim1000$ atoms so that we could perform a thorough analysis of the accuracy and efficiency of SPARC within the computational resources routinely available to us. However, larger systems can indeed be studied, e.g., we have performed a simulation for Si$_{3145}$H$_{876}$  using SPARC. Nevertheless, we note that there is scope for significant improvement in SPARC. Specifically, the subspace eigenvalue problem---currently solved in serial---is expected to become the dominant cost for systems consisting of tens of thousands of electrons. Therefore, incorporating efficient and scalable parallel eigendecomposition techniques into SPARC is currently being undertaken by the authors. These improvements along with optimization of code are expected to further improve the efficiency of SPARC. The extension of SPARC to enable the study of systems with periodicity in one, two, and three directions is also a worthy subject for future work, and is therefore being pursued by the authors.


\section*{Acknowledgements}
\noindent The authors gratefully acknowledge the support of National Science Foundation under Grant Number $1333500$. The authors also gratefully acknowledge the valuable comments and suggestions of the anonymous referees. 


\bibliographystyle{ReferenceStyle}

\newpage

\appendix

\noindent {\LARGE \bf Appendix}

\section{Electrostatic correction for overlapping pseudocharge densities} \label{Appendix:Correct:RepulsiveEnergy}
In ab-initio calculations, even if the pseudopotential approximation is employed, the repulsive energy is still calculated with the nuclei treated as point charges. The electrostatic formulation employed in this work does not make this distinction, resulting in disagreement with convention for overlapping pseudocharge densities. The correction to the repulsive energy which reestablishes agreement can be written as \cite{Suryanarayana2014524}
\begin{eqnarray} 
E_c(\bR) & = & \frac{1}{2} \int_{\R^3} \left( \tilde{b}(\bx,\bR) + b(\bx,\bR) \right) V_c(\bx,\bR) \, \mathrm{d\bx} + \frac{1}{2}\sum_{J=1}^{N} \int_{\R^3} b_J(\bx,\bR_J) V_J(\bx,\bR_J) \, \mathrm{d\bx} \nonumber \\
& & - \frac{1}{2}\sum_{J=1}^{N} \int_{\R^3} \tilde{b}_J(\bx,\bR_J) \tilde{V}_J(\bx,\bR_J) \, \mathrm{d\bx} \,, \label{Eqn:RepulsiveCorrectionEnergy}
\end{eqnarray} 
where 
\begin{equation} \label{Eqn:Vc}
 V_c(\bx,\bR) =  \sum_{J=1}^{N} \left(\tilde{V}_J(\bx,\bR_J)-V_J(\bx,\bR_J)\right) \,.
\end{equation}
In addition, $\tilde{b}$ denotes the reference pseudocharge density, and $\tilde{b}_J$ represents the spherically symmetric and compactly supported reference charge density of the $J^{th}$ nucleus that generates the potential $\tilde{V}_J$, i.e., 
\begin{equation}
\tilde{b}(\bx,\bR)= \sum_{J=1}^{N} \tilde{b}_J(\bx,\bR_J) \,, \quad \tilde{b}_J(\bx,\bR_J)= - \frac{1}{4 \pi} \nabla^2 \tilde{V}_J(\bx,\bR_J) \,, \quad \int_{\R^3} \tilde{b}_J(\bx,\bR_J) \, \mathrm{d\bx} = Z_J \,.
\end{equation}
The discrete form of the repulsive energy correction is obtained by approximating the integrals in Eqn. \ref{Eqn:RepulsiveCorrectionEnergy} using the integration rule in Eqn. \ref{Eqn:IntApprox}:
\begin{equation}
E_c^h = \frac{1}{2} h^3 \sum_{i=1}^{n_1} \sum_{j=1}^{n_2} \sum_{k=1}^{n_3} \left( (\tilde{b}^{(i,j,k)} + b^{(i,j,k)}) V_c^{(i,j,k)} + \sum_{J=1}^{N} b_J^{(i,j,k)} V_J^{(i,j,k)} - \sum_{J=1}^{N} \tilde{b}_J^{(i,j,k)} \tilde{V}_{J}^{(i,j,k)}  \right) \,, \label{Eqn:Ec:Disc}
\end{equation}
where $V_c^{(i,j,k)}$ is obtained using Eqn. \ref{Eqn:Vc}. 

The correction in the atomic forces arising from the overlapping pseudocharges can be written as \cite{Suryanarayana2014524}
\begin{eqnarray}
\mathbf{f}_{J,c}(\bR) & = & \frac{1}{2} \int_{\R^3} \bigg[ \nabla \tilde{b}_{J}(\bx,\bR_{J}) \left(V_c(\bx,\bR)- \tilde{V}_{J}(\bx,\bR_{J})\right) + \nabla b_{J}(\bx,\bR_{J}) \left(V_c(\bx,\bR)+V_{J}(\bx,\bR_{J})\right) \nonumber \\ 
& + & \left(\nabla \tilde{V}_{J}(\bx,\bR_{J}) - \nabla V_{J}(\bx,\bR_{J})\right) \left(\tilde{b}(\bx,\bR)+b(\bx,\bR)\right) + b_{J}(\bx,\bR_{J}) \nabla V_{J}(\bx,\bR_{J}) \label{Eqn:RepulsiveCorrectionForce} \\ 
& - & \tilde{b}_{J}(\bx,\bR_{J}) \nabla \tilde{V}_{J}(\bx,\bR_{J}) \bigg] \,\mathrm{d\bx} \,,  \nonumber
\end{eqnarray}
whose discrete form is
\begin{eqnarray}
\mathbf{f}_{J,c}^h & = & \frac{1}{2} h^3  \sum_{i=1}^{n_1} \sum_{j=1}^{n_2} \sum_{k=1}^{n_3} \bigg( \nabla_h \tilde{b}\big|^{(i,j,k)} \left(V_{c}^{(i,j,k)} - \tilde{V}_{J}^{(i,j,k)}\right) +  \nabla_h b_{J}\big|^{(i,j,k)} \left(V_{c}^{(i,j,k)}+  V_{J}^{(i,j,k)}\right)  \nonumber \\ 
& & + \nabla_h (\tilde{V}_{J} - V_{J})\big|^{(i,j,k)} \left(\tilde{b}^{(i,j,k)}+b^{(i,j,k)}\right) + b_{J}^{(i,j,k)} \nabla_h V_{J}\big|^{(i,j,k)} - \tilde{b}_{J}^{(i,j,k)} \nabla_h \tilde{V}_{J}\big|^{(i,j,k)}  \bigg) \,. \label{Eqn:LF:Disc} 
\end{eqnarray}
It is worth noting that an alternative to the above formulation is to correct for the error in the repulsive energy and the corresponding force by only considering the pseudocharges that overlap. However, this requires the creation of neighbor lists, which need to be updated at every relaxation step. In SPARC, we employ the corrections in Eqns. \ref{Eqn:Ec:Disc} and \ref{Eqn:LF:Disc} because of their simplicity and accuracy in the context of our electrostatic formulation, and their efficiency in the setting of scalable high performance computing. 

\begin{figure}[H]
\centering
\subfloat[Energy correction]{\label{Fig:energy_correction}\includegraphics[keepaspectratio=true,width=0.46\textwidth]{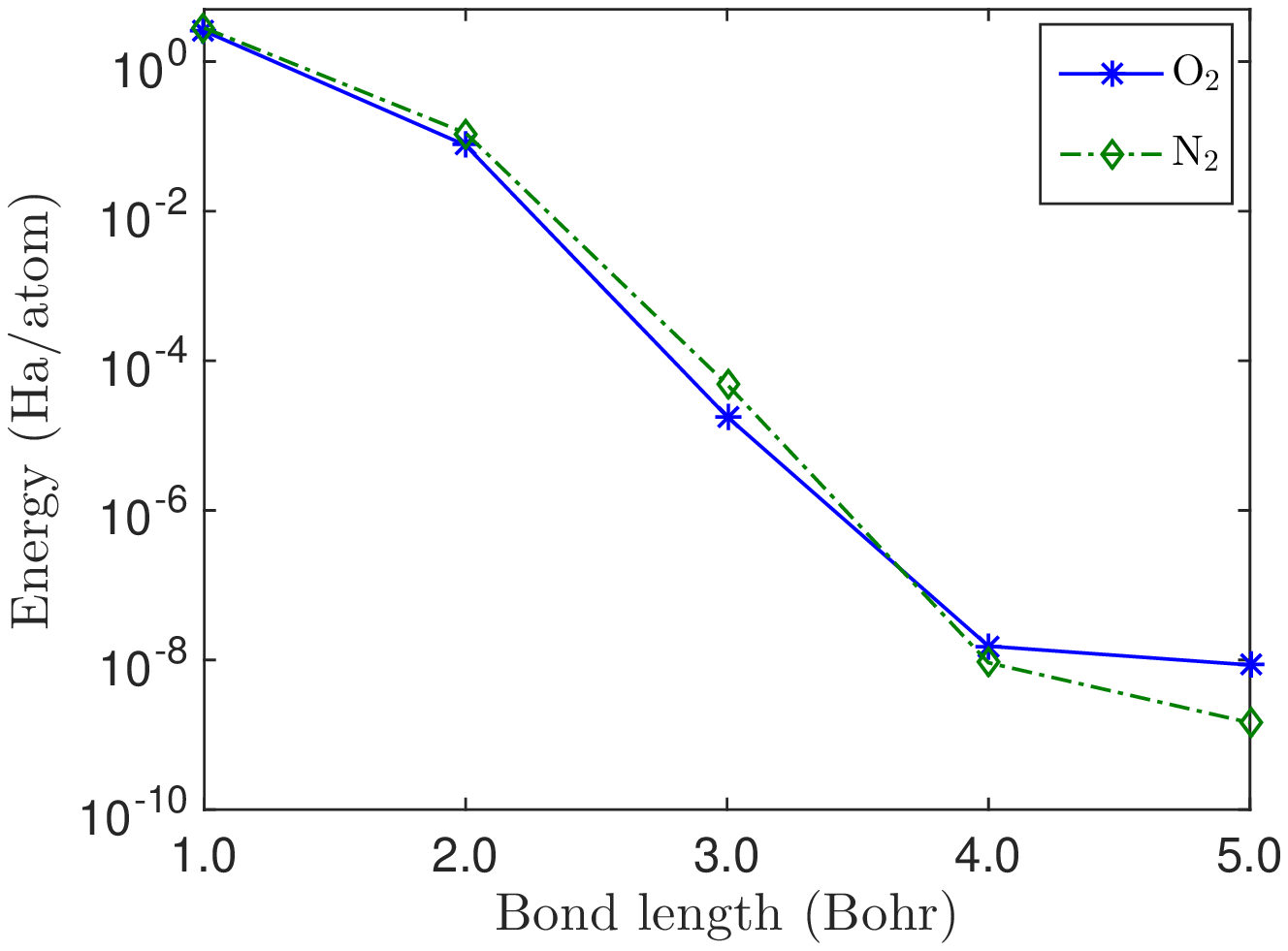}}
\subfloat[Atomic force correction]{\label{Fig:force_correction}\includegraphics[keepaspectratio=true,width=0.46\textwidth]{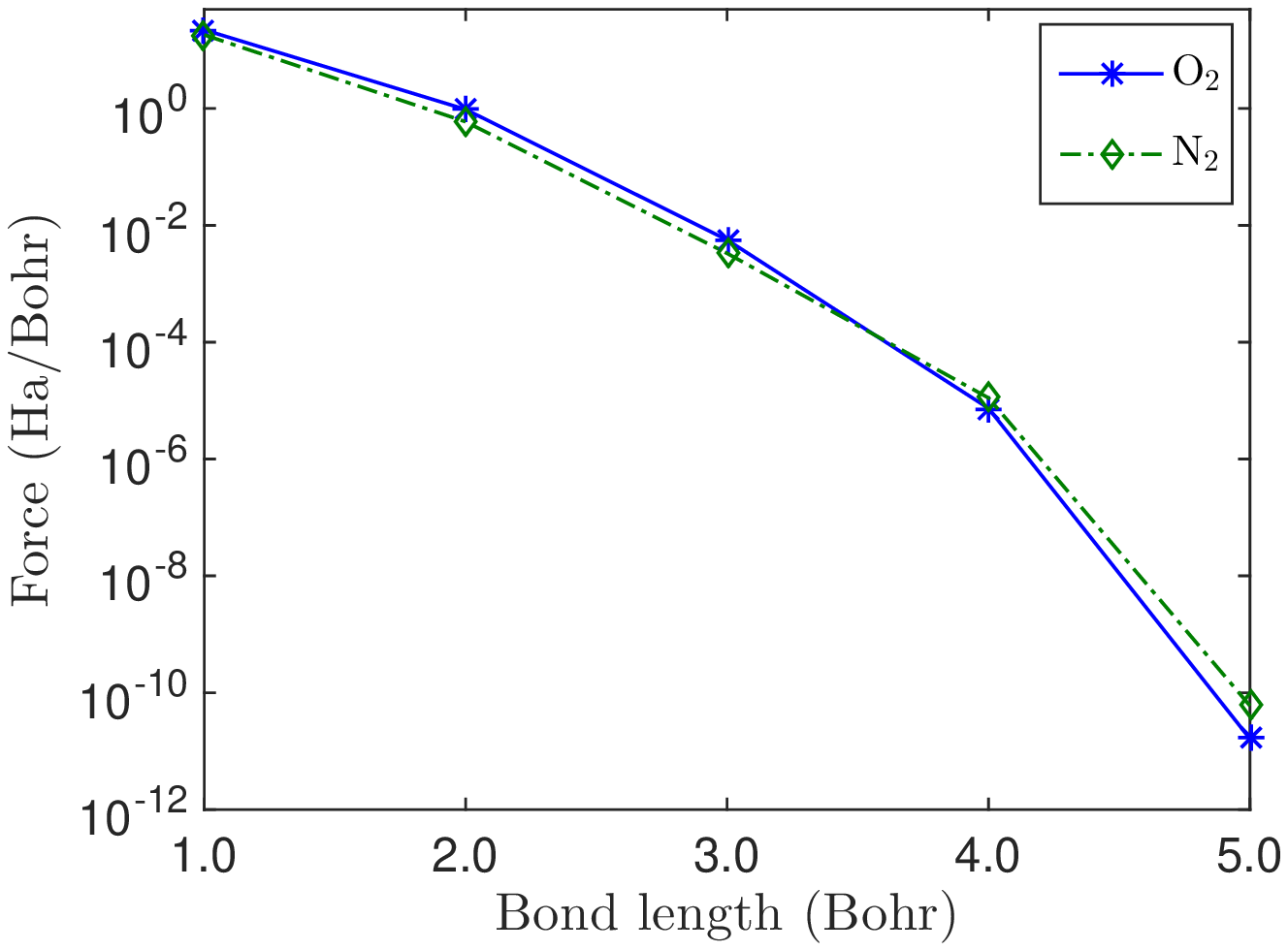}}
\caption{Variation in magnitude of the repulsive energy and corresponding atomic force corrections as a function of interatomic distance for the N$_2$ and O$_2$ molecules.}
\label{Fig:Correction}
\end{figure}

In order to demonstrate the importance of the aforedescribed energy and force corrections, we plot their values as a function of interatomic distance for the N$_2$ and O$_2$ molecules ($h=0.2$ Bohr) in Fig.~\ref{Fig:Correction}. We observe that though the magnitude of the energy and force corrections reduce as the distance between the atoms is increased, they are still significant at the equilibrium bond length. Notably, even at distances of around $2 r_J^c \sim 3$ Bohr (Table \ref{Table:CutoffRadiiPseudopotential}), the force corrections have magnitude of approximately $0.01$ Ha/Bohr. This is because, even though the Troullier-Martins pseudopotentials have non-local projectors that are identically zero outside $r_J^c$, each of pseudopotentials individually approach the Coulomb potential at values that are noticeably larger than $r_J^c$. This highlights the need for incorporating the repulsive energy and corresponding atomic force corrections within SPARC, and possibly other real-space DFT implementations that utilize the reformulation of the electrostatics in terms of the pseudocharges, particularly for structural relaxations and molecular dynamics simulations.

\section{Pseudopotential parameters} \label{Appendix:Pseudopotential}
In Table \ref{Table:CutoffRadiiPseudopotential}, we list the cutoff radii ($r_J^c$) used for generating the different angular momentum components within the Troullier-Martins pseudopotential. We choose the $l=0$ pseudopotential component as local in all the simulations.

\begin{table}[H]
\centering
\begin{tabular}{ccccc}
\hline 

\multirow{2}{*}{Atom type} & \multicolumn{3}{c}{Radial cutoff (Bohr)}  \\
          &  $l=0$        & $l=1$        & $l=2$         \\
\hline
H         & $1.25$         &  $-$        & $-$               \\
C         & $1.50$         &  $1.54$        & $-$               \\
N        & $1.50$         &  $1.50$        & $-$              \\
O         & $1.45$         &  $1.45$        & $-$            \\
Si        & $1.80$         &  $1.80$        & $1.80$            \\
Pt         & $2.45$         &  $2.45$        & $2.45$               \\
Au         & $2.60$         &  $2.60$        & $2.60$            \\
\hline
\end{tabular}
\caption{Cutoff radii for non-local projectors within the Troullier-Martins pseudopotential.}
\label{Table:CutoffRadiiPseudopotential}
\end{table}


\section{Properties of the discrete pseudocharge density} \label{Appendix:Pseudocharge}
\begin{figure}[H]
\centering
\includegraphics[keepaspectratio=true,width=0.46\textwidth]{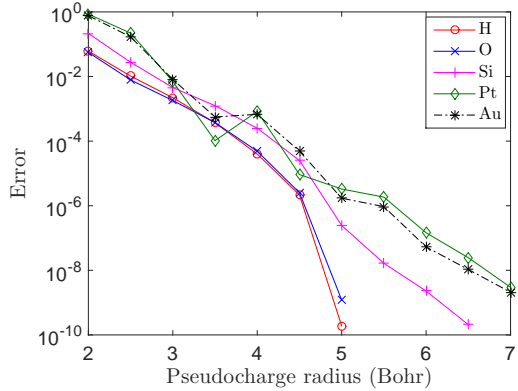}
\caption{Normalized error in the net enclosed charge as a function of pseudocharge radius. The results for carbon and nitrogen are identical to oxygen.}
\label{Fig:PseudochargeDecay} 
\end{figure}

The continuous pseudocharge density for the atom positioned at $\bR_J$ has compact support in a sphere of radius $r_J^c$ centered at $\bR_J$, where $r_J^c$ is the cutoff radius for the local component of the pseudopotential. Though the corresponding discrete pseudocharge density has infinite extent, it still possesses exponential decay. This is evident from  Fig. \ref{Fig:PseudochargeDecay}, where we plot the normalized error in the net enclosed charge as a function of the pseudocharge radius $r_J^b$ for a mesh-size of $h=0.5$ Bohr. It is clear that a suitable finite truncation radius can indeed be chosen such that there is no significant loss of accuracy.

\begin{figure}[H]
\centering
\includegraphics[keepaspectratio=true,width=0.46\textwidth]{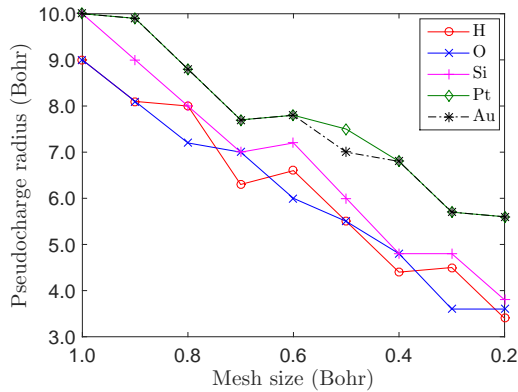}
\caption{Variation of pseudocharge radius as a function of mesh spacing. The results for carbon and nitrogen are identical to oxygen.}
\label{Fig:PseudochargeRadii} 
\end{figure}

In this work, we choose the truncation radius  $r_J^b$ for each pseudocharge density such that Eqn. \ref{Eqn:PseudochargeRadiusChoice} is satisfied to within a tolerance of $\varepsilon_b= 10^{-8}$. In Fig. \ref{Fig:PseudochargeRadii}, we plot the $r_J^b$ required to achieve this desired accuracy as a function of mesh-size. It is clear that as the mesh becomes finer, $r_J^b$ becomes smaller, with $r_J^b \rightarrow r_J^c$ as $h \rightarrow 0$. The slight non-monotonicity of the curves plotted in Fig. \ref{Fig:PseudochargeRadii} is due to the fact that $r_J^b$ is chosen to be a multiple of the mesh size $h$ in SPARC. 

\section{Comparison of SPARC with other finite-difference DFT codes} \label{Appendix:RealSpaceComparison}
\begin{figure}[H]
\centering
\subfloat[Energy]{\label{fig:energyConvergence_rs}\includegraphics[keepaspectratio=true,width=0.46\textwidth]{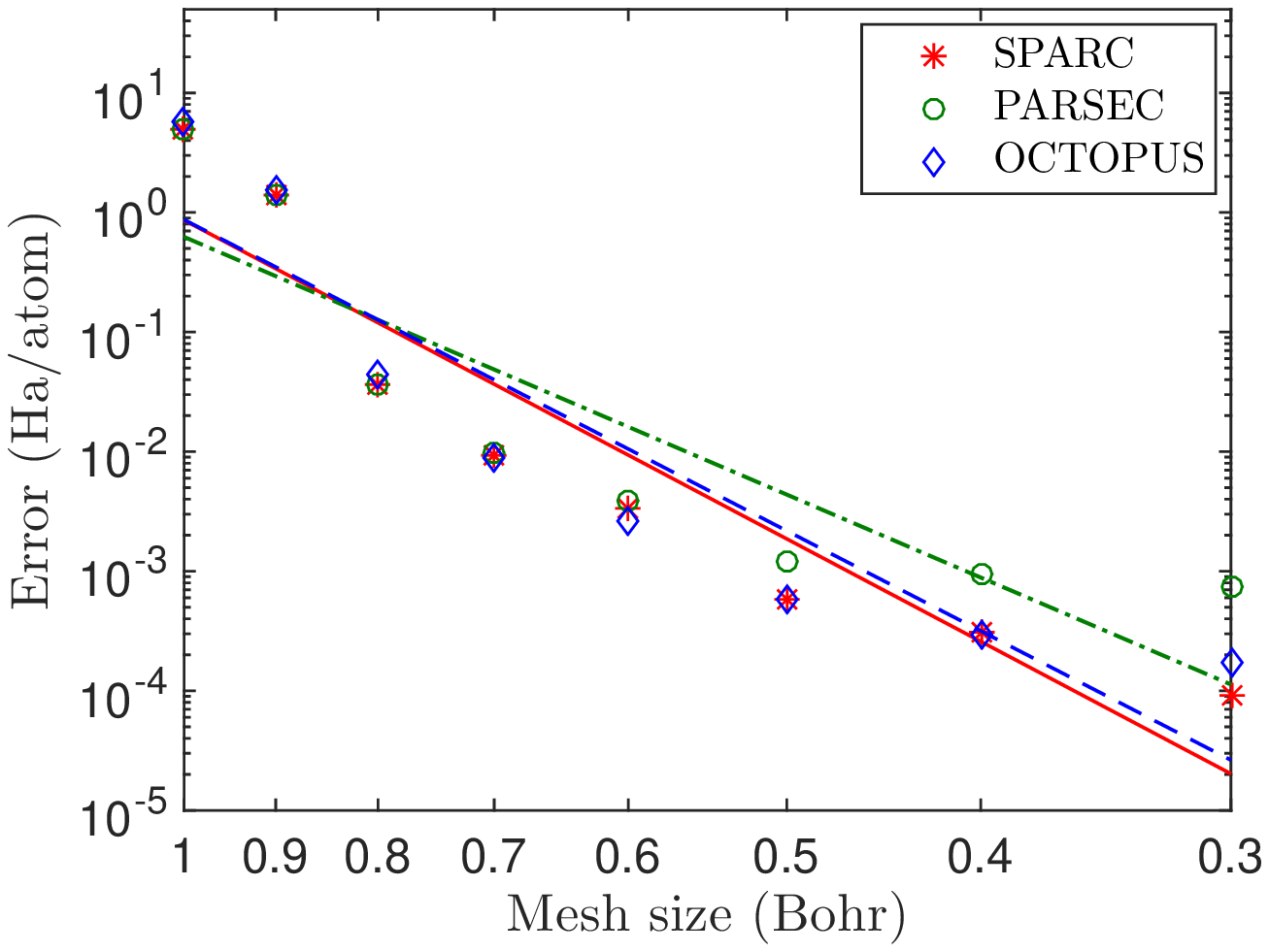}}
\subfloat[Forces]{\label{fig:forceConvergence_rs}\includegraphics[keepaspectratio=true,width=0.46\textwidth]{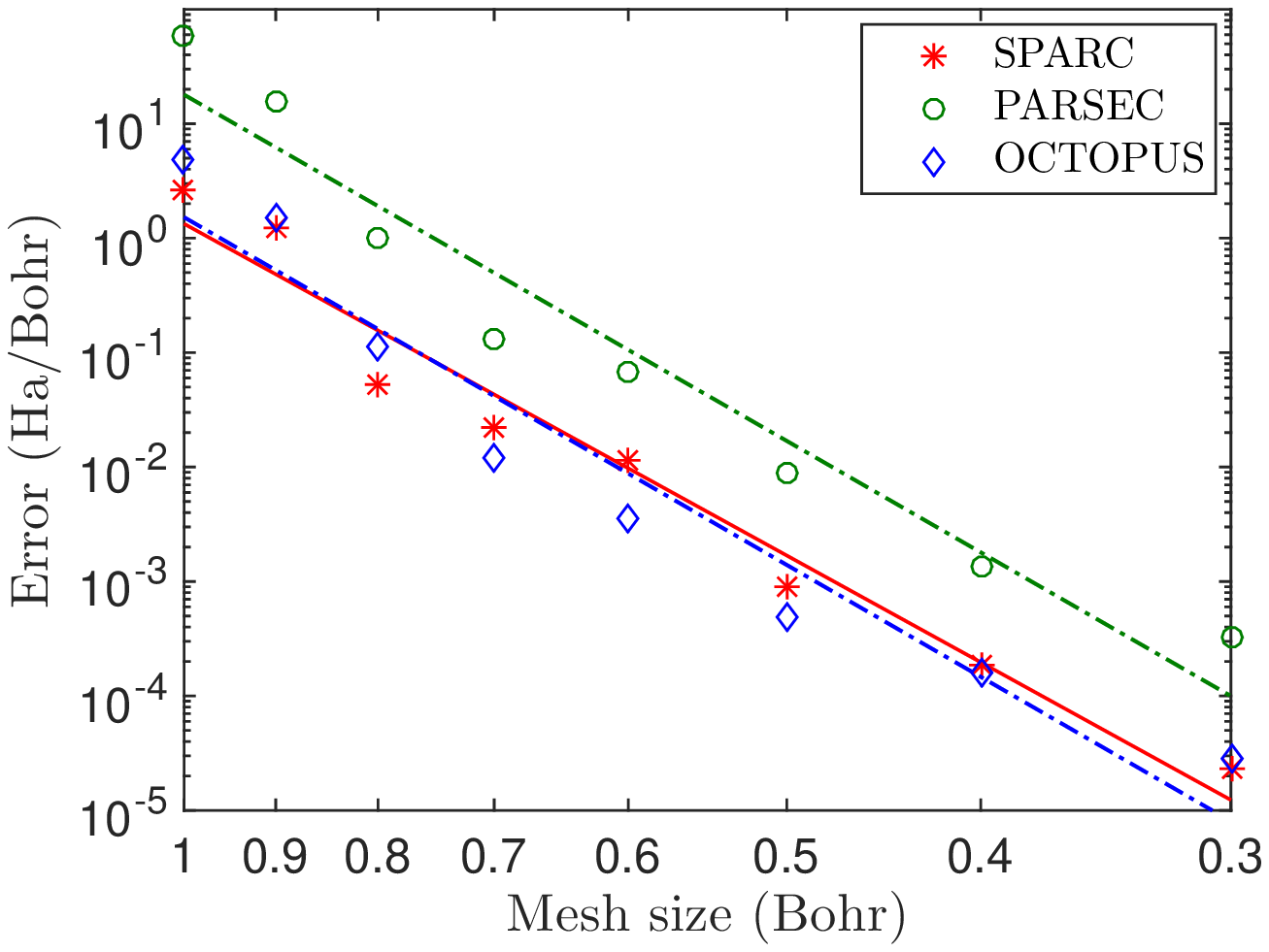}}
\caption{Convergence of the energy and atomic forces with respect to mesh size to reference planewave result for the Si$_{29}$H$_{36}$ cluster.}
\label{fig:convergenceDiscretization_rs}
\end{figure}

We now study the performance of SPARC relative to PARSEC \cite{chelikowsky1994finite} and OCTOPUS \cite{OCTOPUS}, two well-established DFT codes that employ the finite-difference discretization. First, we determine the convergence in energy and atomic forces as a function of mesh-size $h$ for the Si$_{29}$H$_{36}$ cluster with the central Silicon atom perturbed by [$0.4$ $0.3$ $0.6$] Bohr. All errors are defined with respect to ABINIT, wherein we employ a plane-wave cutoff of $30$ Ha and domain size of $\{L_1,L_2,L_3\} =\{42,42,42\}$ Bohr, which results in reference energies and forces that are converged to within $5.0 \times 10^{-6}$ Ha/atom and $5.0 \times 10^{-6}$ Ha/Bohr, respectively. On performing a fit to the data presented in Fig. \ref{fig:convergenceDiscretization_rs}, we obtain $\mathcal{O}(h^{8.85})$, $\mathcal{O}(h^{7.15})$, and $\mathcal{O}(h^{8.64})$ convergence in energy for SPARC, PARSEC, and OCTOPUS, respectively. Correspondingly, we obtain $\mathcal{O}(h^{9.62})$, $\mathcal{O}(h^{10.05})$, and $\mathcal{O}(h^{10.09})$ convergence in the forces. Though the convergence rates of all three codes are comparable, the associated prefactor in PARSEC is noticeably larger, particularly for the atomic forces\footnote{In recent work, the quality of the atomic forces in PARSEC  has been improved by using high-order integrations \cite{BobSchChe15}.}. 

\begin{figure}[H]
\centering
\subfloat[Strong scaling]{\label{Fig:RealSpaceStrongScaling}\includegraphics[keepaspectratio=true,width=0.46\textwidth]{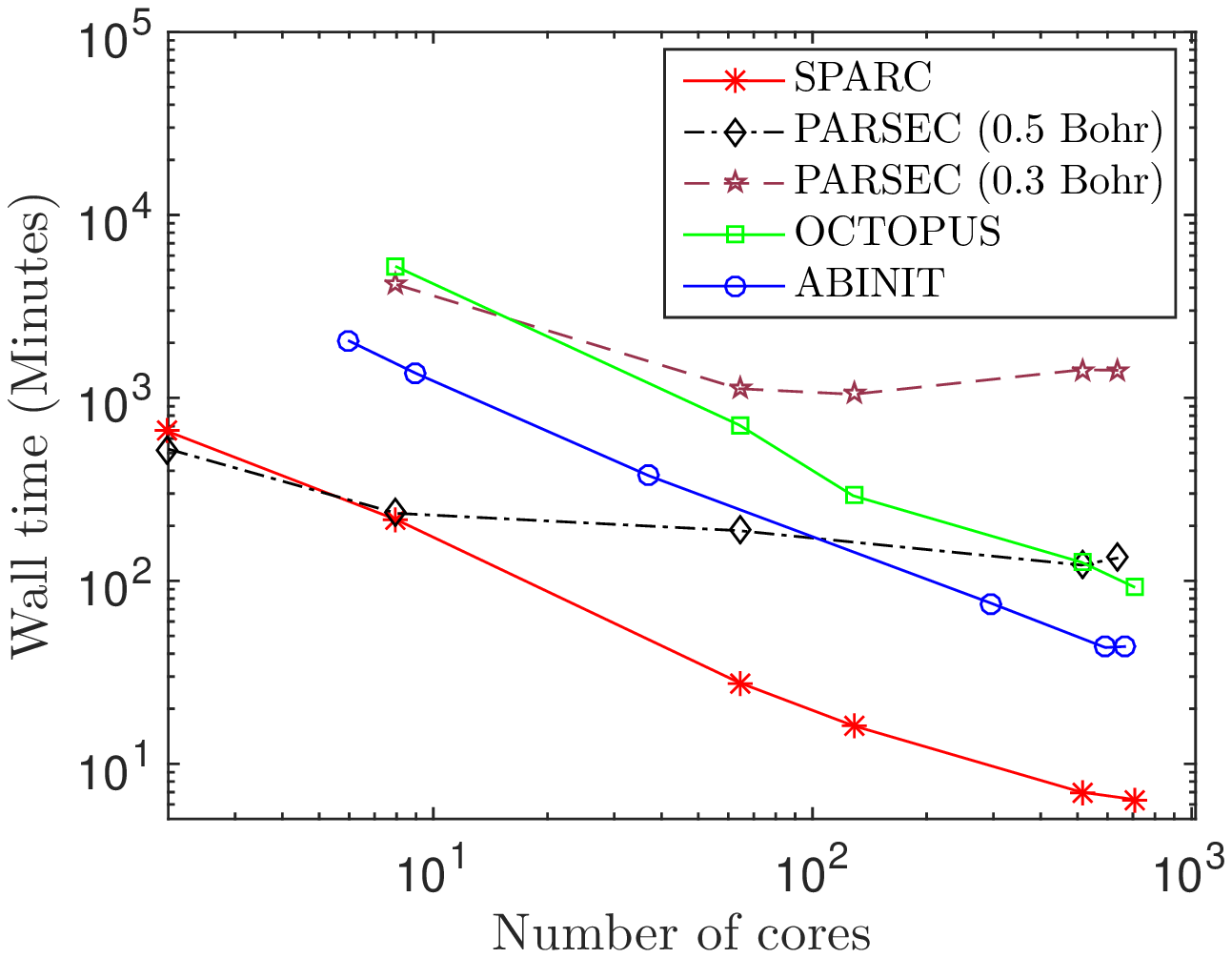}}
\subfloat[Weak scaling]{\label{Fig:RealSpaceWeakScaling}\includegraphics[keepaspectratio=true,width=0.46\textwidth]{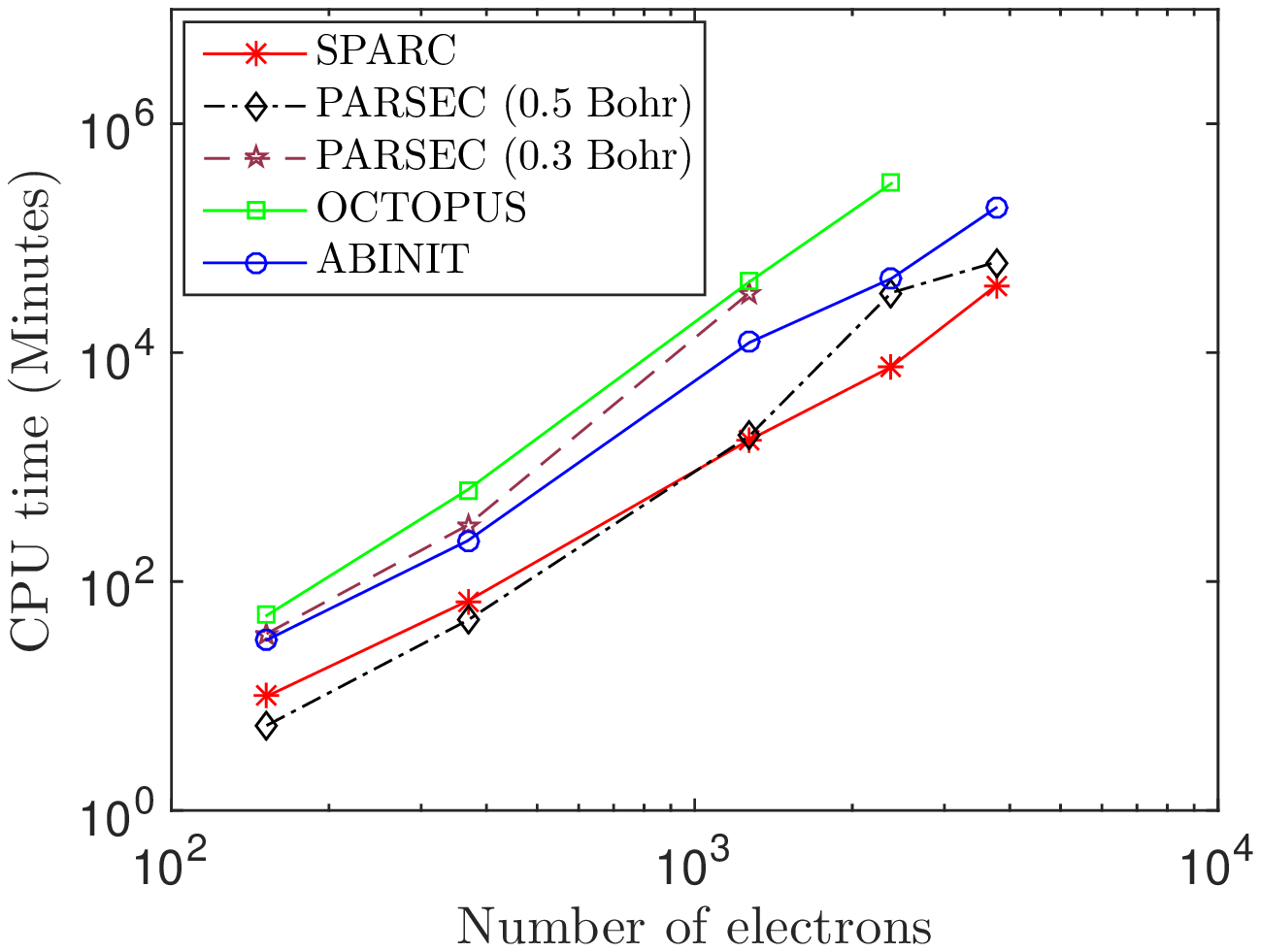}}
\caption{Strong and weak scaling behavior for hydrogen passivated silicon nanoclusters. The system utilized for strong scaling is Si$_{275}$H$_{172}$. The systems employed for weak scaling are Si$_{29}$H$_{36}$, Si$_{71}$H$_{84}$, Si$_{275}$H$_{172}$, Si$_{525}$H$_{276}$ and Si$_{849}$H$_{372}$. The time taken for the first SCF iteration has been excluded.}
\label{Fig:StrongWeakScaling:RealSpace}
\end{figure}

Next, we perform the strong and weak scaling tests described in Section \ref{Subsec:Scaling} for SPARC, PARSEC, OCTOPUS, and ABINIT. In order to achieve the desired accuracy of $0.001$ Ha/atom in the energy and $0.001$ Ha/Bohr in the atomic forces, we choose a plane-wave cutoff of $16$ Ha in ABINIT, and mesh-sizes of $h=0.5$ Bohr, $h=0.3$ Bohr, and $h=0.5$ Bohr in SPARC, PARSEC, and OCTOPUS, respectively. We also determine the timings for $h=0.5$ Bohr in PARSEC. We present the results so obtained in Fig. \ref{Fig:StrongWeakScaling:RealSpace}, wherein the time taken for the first SCF iteration has been excluded.\footnote{The version of PARSEC used in this study employs diagonalization in the first SCF iteration, which can be particularly expensive. This can be overcome using the technique that has been recently proposed by some of the PARSEC developers \cite{zhou2014chebyshev}, which is also a part of the SPARC formulation and implementation.} In strong scaling, the minimum wall time achieved by SPARC is smaller by factors of $21$ ($221$ for $h=0.3$ Bohr in PARSEC), $14$, and $6.8$ relative to PARSEC, OCTOPUS, and ABINIT, respectively. In weak scaling, the increase in CPU time with number of electrons for SPARC, PARSEC, OCTOPUS, and ABINIT is $\mathcal{O}(N_e^{2.54})$, $\mathcal{O}(N_e^{3})$, $\mathcal{O}(N_e^{3.19})$, and $\mathcal{O}(N_e^{2.75})$, respectively. It is clear that SPARC is able to outperform PARSEC because of the significantly higher efficiency in strong scaling. In OCTOPUS, $>70 \%$ (increases with system size) of the time is spent in the function \texttt{GRAM\_SCHMIDT} (orthogonalizations), which suggests that CheFSI is significantly superior to Conjugate Gradients (default eigensolver in OCTOPUS) for real-space DFT calculations. Notably, even when RMM-DIIS is employed in OCTOPUS, SPARC demonstrates superior performance. For example, SPARC is faster than OCTOPUS by factors of $3.9$ and $5.76$ for the Si$_{275}$H$_{172}$ and Si$_{525}$H$_{276}$ systems in the weak scaling study, respectively. This is a consequence of the significantly larger number of iterations required by RMM-DIIS in OCTOPUS.

Finally, we compare the minimum wall time---excluding the time for the first SCF iteration---that can be achieved by SPARC, PARSEC, OCTOPUS, and ABINIT for the aforementioned hydrogen passivated nanoclusters. From the results presented in Table \ref{Table:realspaceTime}, we observe that SPARC demonstrates speedup by up to factors of $21$ ($164$ for $h=0.3$ Bohr in PARSEC), $15$, and $7$ compared to PARSEC, OCTOPUS, and ABINIT, respectively. Overall, these results demonstrate that SPARC is an efficient DFT formulation and implementation that is highly competitive with well-established finite-difference and plane-wave codes. In addition, previous finite-difference DFT codes are unable to consistently outperform plane-wave codes in achieving the desired accuracy for the examples considered here. 

\begin{table}[H]
\centering
\begin{tabular}{cccccc}
\hline
\multirow{2}{*}{System} &   SPARC     & PARSEC & PARSEC &OCTOPUS & ABINIT  \\  
       & $h=0.5$ Bohr & $h=0.5$ Bohr & $h=0.3$ Bohr & $h=0.5$ Bohr & $E_{cut}=16$ Ha\\
\hline
Si$_{29}$H$_{36}$        &   $0.55$ $(128)$  & $3.22$ $(8)$ & $18.92$ $(16)$ & $2.8$ $(256)$ & $6.25$ $(106)$     \\
Si$_{71}$H$_{84}$        &   $0.96$ $(320)$  & $10.32$ $(64)$ & $50.53$ $(64)$ & $10.15$ $(512)$ & $6.70$ $(321)$    \\
Si$_{275}$H$_{172}$      &   $6.39$ $(704)$  & $121.96$ $(512)$ & $1046.00$ $(128)$ & $92.78$ $(1024)$ & $43.76$ $(666)$    \\
Si$_{525}$H$_{276}$      &   $29.00$ $(960)$ & $619.72$ $(512)$ & --- & $366.12$ $(1024)$ & $203.30$ $(1008)$ \\
\hline
\end{tabular}
\caption{Minimum wall time in minutes for hydrogen passivated silicon nanoclusters. The number in brackets represents the number of cores on which the minimum wall time is achieved. The time taken for the first SCF iteration has been excluded.}
\label{Table:realspaceTime}
\end{table}


\end{document}